\documentclass[12pt]{amsart}
\usepackage{graphicx}
\usepackage{amsthm}
\usepackage{amsmath}
\usepackage{amssymb}
\usepackage{lscape}
\usepackage{multicol}
\usepackage{multirow}
\usepackage{amsmath}
\usepackage{amsfonts}
\usepackage{natbib}

\usepackage{tikz}
\usetikzlibrary{arrows,positioning} 
\usepackage[section]{placeins}

\newtheorem{theorem}{Theorem}[section]
\newtheorem{example}[theorem]{Example}

\newfam\bboardfam
\font\tenbboard=msbm10  
 \font\sevenbboard=msbm7
   \font\fivebboard=msbm5 
\textfont\bboardfam=\tenbboard
\scriptfont\bboardfam=\sevenbboard
\scriptscriptfont\bboardfam=\fivebboard
\def\bboard{\fam\bboardfam\tenbboard}
\def\R{{\bboard R}}     % Real numbers
\newcommand{\Rpl}{\mbox{$\R_{> 0}$}}

\DeclareMathOperator{\sign}{sign}
\DeclareMathOperator{\diag}{diag}
\begin{document}

\title[MAPK's and multistationarity due to toric steady states]{MAPK's networks and their 
capacity for multistationarity due to toric steady states}

\author{Mercedes P\'erez Mill\'an and Adri\'an G. Turjanski}
\address{MPM: Dto.\ de Matem\'atica, FCEN, Universidad de Buenos Aires, 
Ciudad Universitaria, Pab.\ I, C1428EGA Buenos Aires, Argentina;
Dto. de Ciencias Exactas, CBC, Universidad de Buenos Aires, Ramos 
Mej\'{i}a 841, C1405CAE Buenos Aires, Argentina. 
AGT:Dto.\ de Qu\'imica Biol\'ogica, FCEN, Universidad de Buenos Aires, 
Ciudad Universitaria, Pab.\ II, C1428EGA Buenos Aires, Argentina.}
\email{mpmillan@dm.uba.ar,adrian@qi.fcen.uba.ar}

\begin{abstract}
Mitogen-activated protein kinase (MAPK) signaling pathways play an essential role in the 
transduction of environmental stimuli to the nucleus, thereby regulating a variety of 
cellular processes, including cell proliferation, differentiation and programmed cell 
death. The components of the MAPK extracellular activated protein kinase (ERK) cascade 
represent attractive targets for cancer therapy as their aberrant activation is a frequent 
event among highly prevalent human cancers. MAPK networks are a model for computational simulation,
mostly using Ordinary and Partial Differential Equations. Key results showed that these networks can have 
switch-like behavior, bistability and oscillations. 
In this work, we consider three representative ERK networks, one with a negative 
feedback loop, which present a binomial steady state ideal under mass-action kinetics.
We therefore apply the theoretical result present in \cite{psc11} to find 
a set of rate constants that allow two significantly different stable 
steady states in the same stoichiometric compatibility class for each network. 
Our approach makes it possible to study certain aspects of the system, such as multistationarity, 
without relying on simulation, since we do not assume a priori any constant but the topology of the network.
As the performed analysis is general it could be applied to many other important biochemical networks.

  \vskip 0.1cm
% : Keywords

\end{abstract}

\maketitle

\section{INTRODUCTION}\label{sec:intro}

Mitogen-activated protein kinases (MAPKs) are serine/threonine kinases that play an
essential role in signal transduction by modulating gene transcription in the nucleus in 
response to changes in the cellular environment. MAPKs participate in a number of 
disease states including chronic inflammation and cancer \citep{davis,kyriakis,pearson,
schaeffer,zarubin} as they control key cellular functions, including differentiation, 
proliferation, migration and apoptosis. In humans, there are several members of the 
MAPK superfamily which can be divided in groups as each group can be stimulated by 
a separate protein kinase cascade that includes the sequential activation of a 
specific MAPK kinase kinase (MAPKKK) and a MAPK kinase (MAPKK), which in turn 
phosphorylates and activates their downstream MAPKs \citep{pearson,tvg07}. 
These signaling modules 
have been conserved throughout evolution, from plants, fungi, nematodes, insects, 
to mammals \citep{widmann}. Among the MAPK pathways, the mechanisms governing the 
activation of ERK2 have been the most extensively studied, the MAPKK is MEK2 and the 
MAPKKK is RAF which can be activated by RAS. Impeding the function of ERK2 
prevents cell proliferation in response to a variety of growth factors \citep{pages} 
and its overactivity is sufficient to transform cells in culture \citep{mansour}. 
RAS, RAF and MEK2 have been intensively studied for the development of 
cancer inhibitors with several of them in the market. Indeed, the wealth of available 
cellular and biochemical  information on the nature of the signaling routes that activate MAPK 
has enabled the use of computational approaches to study MAPK activation, 
thus becoming a prototype for systems biology studies \citep{hornberg,schoeberl}.

In the present work, we study the capacity for multistationarity of three systems which 
involve the activation of a MAPKKK then a MAPKK and finally a MAPK and are of general
application but have been proposed previously for the extracellular 
signal-regulated kinase (ERK) cascade: The first network 
is the most frequent in the literature \citep{kholo00,ferrell96} and is the simple 
sequential activation. The second one differs from the first one in the phosphatases, 
which we assume to be equal for the last two layers of the cascades \citep{fujioka}.
The third network includes a negative feedback between pRAF and ppERK in which the latter 
acts as a kinase for the former, producing a new phosphorylated and inactive form Z 
\citep{al01,dough05,fritsche}.

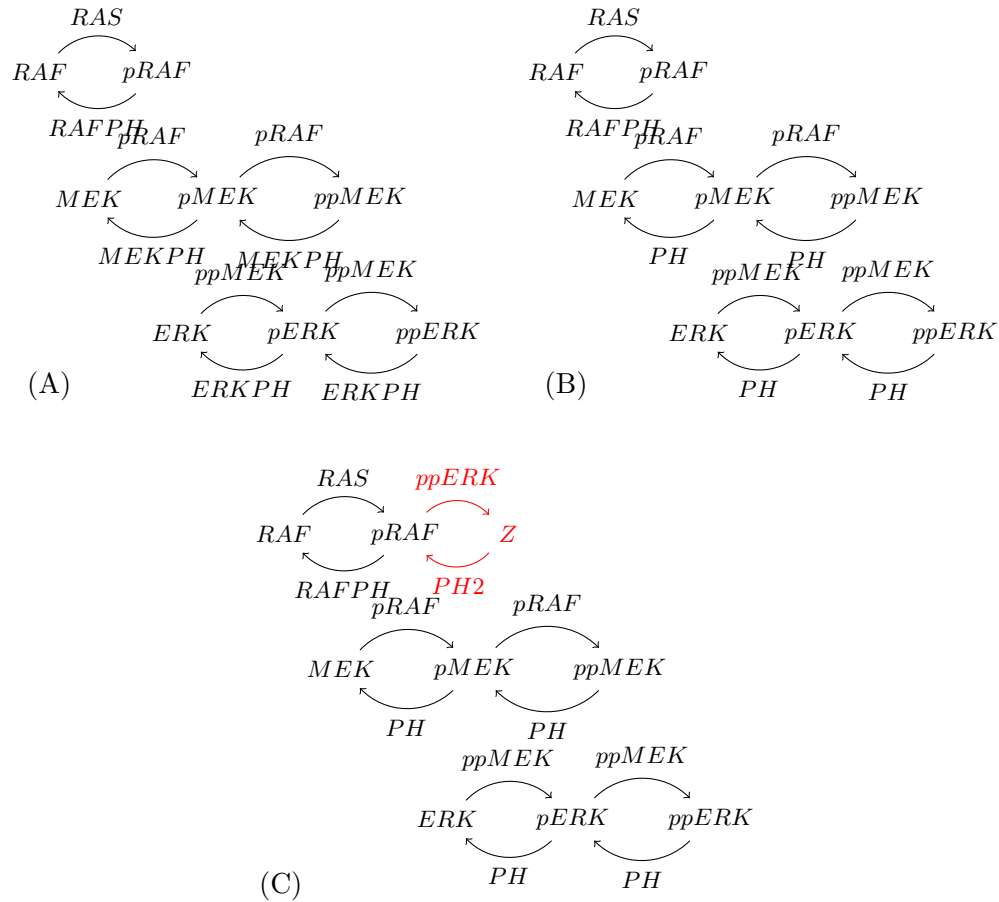
\begin{figure}[!hb]
\begin{tabular}{cc}
 \begin{tikzpicture}[node distance=0.45cm,scale=0.35]
 \node[] at (-1,0) (dummy) {};
 \node[left=of dummy] (s0) {\scriptsize $RAF$};
  \node[right=of s0] (s1) {\scriptsize $pRAF$}
   edge[->, bend left=45] node[below] {\scriptsize $RAFPH$} (s0)
   edge[<-, bend right=45] node[above] {\scriptsize $RAS$} (s0);
 \node[] at (1.1,-4.8) (dummy2) {};
 \node[left=of dummy2] (p0) {\scriptsize $MEK$};
 \node[right=of p0] (p1) {\scriptsize $pMEK$}
   edge[->, bend left=45] node[below] {\scriptsize $MEKPH$} (p0)
   edge[<-, bend right=45] node[above] {\scriptsize $pRAF$} (p0); 
 \node[right=of p1] (p2) {\scriptsize $ppMEK$}
   edge[->, bend left=45] node[below] {\scriptsize $MEKPH$} (p1)
   edge[<-, bend right=45] node[above] {\scriptsize $pRAF$} (p1); 
 \node[] at (4.5,-9.9) (dummy3) {};
 \node[left=of dummy3] (r0) {\scriptsize $ERK$};
 \node[right=of r0] (r1) {\scriptsize $pERK$}
   edge[->, bend left=45] node[below] {\scriptsize $ERKPH$} (r0)
   edge[<-, bend right=45] node[above] {\scriptsize $ppMEK$} (r0); 
 \node[right=of r1] (r2) {\scriptsize $ppERK$}
   edge[->, bend left=45] node[below] {\scriptsize $ERKPH$} (r1)
   edge[<-, bend right=45] node[above] {\scriptsize $ppMEK$} (r1); 
 \node[] at (-3.8,-11.9) (dummy3) {\small(A)};
\end{tikzpicture} &
 \begin{tikzpicture}[node distance=0.45cm,scale=0.35]
 \node[] at (-1,0) (dummy) {};
 \node[left=of dummy] (s0) {\scriptsize $RAF$};
  \node[right=of s0] (s1) {\scriptsize $pRAF$}
   edge[->, bend left=45] node[below] {\scriptsize $RAFPH$} (s0)
   edge[<-, bend right=45] node[above] {\scriptsize $RAS$} (s0);
 \node[] at (1.1,-4.8) (dummy2) {};
 \node[left=of dummy2] (p0) {\scriptsize $MEK$};
 \node[right=of p0] (p1) {\scriptsize $pMEK$}
   edge[->, bend left=45] node[below] {\scriptsize $PH$} (p0)
   edge[<-, bend right=45] node[above] {\scriptsize $pRAF$} (p0); 
 \node[right=of p1] (p2) {\scriptsize $ppMEK$}
   edge[->, bend left=45] node[below] {\scriptsize $PH$} (p1)
   edge[<-, bend right=45] node[above] {\scriptsize $pRAF$} (p1); 
 \node[] at (4.5,-9.9) (dummy3) {};
 \node[left=of dummy3] (r0) {\scriptsize $ERK$};
 \node[right=of r0] (r1) {\scriptsize $pERK$}
   edge[->, bend left=45] node[below] {\scriptsize $PH$} (r0)
   edge[<-, bend right=45] node[above] {\scriptsize $ppMEK$} (r0); 
 \node[right=of r1] (r2) {\scriptsize $ppERK$}
   edge[->, bend left=45] node[below] {\scriptsize $PH$} (r1)
   edge[<-, bend right=45] node[above] {\scriptsize $ppMEK$} (r1); 
 \node[] at (-3.8,-11.9) (dummy3) {\small(B)};
\end{tikzpicture} \\ \\
\multicolumn{2}{c}{\begin{tikzpicture}[node distance=0.5cm,scale=0.4]
 \node[] at (-1,0) (dummy) {};
 \node[left=of dummy] (s0) {\scriptsize $RAF$};
  \node[right=of s0] (s1) {\scriptsize $pRAF$}
   edge[->, bend left=45] node[below] {\scriptsize $RAFPH$} (s0)
   edge[<-, bend right=45] node[above] {\scriptsize $RAS$} (s0);
  \node[right=of s1] (s2) {\textcolor{red}{\scriptsize $Z$}}
   edge[->, bend left=45,color=red] node[below] {\scriptsize $PH2$} (s1)
   edge[<-, bend right=45,color=red] node[above] {\scriptsize $ppERK$} (s1); 
 \node[] at (1.1,-4.5) (dummy2) {};
 \node[left=of dummy2] (p0) {\scriptsize $MEK$};
 \node[right=of p0] (p1) {\scriptsize $pMEK$}
   edge[->, bend left=45] node[below] {\scriptsize $PH$} (p0)
   edge[<-, bend right=45] node[above] {\scriptsize $pRAF$} (p0); 
 \node[right=of p1] (p2) {\scriptsize $ppMEK$}
   edge[->, bend left=45] node[below] {\scriptsize $PH$} (p1)
   edge[<-, bend right=45] node[above] {\scriptsize $pRAF$} (p1); 
 \node[] at (4.5,-9.5) (dummy3) {};
 \node[left=of dummy3] (r0) {\scriptsize $ERK$};
 \node[right=of r0] (r1) {\scriptsize $pERK$}
   edge[->, bend left=45] node[below] {\scriptsize $PH$} (r0)
   edge[<-, bend right=45] node[above] {\scriptsize $ppMEK$} (r0); 
 \node[right=of r1] (r2) {\scriptsize $ppERK$}
   edge[->, bend left=45] node[below] {\scriptsize $PH$} (r1)
   edge[<-, bend right=45] node[above] {\scriptsize $ppMEK$} (r1); 
 \node[] at (-4,-11.7) (dummy3) {\small(C)};
\end{tikzpicture}} 
\end{tabular}

 \caption{(A)The ERK network with sequential activation. (B)The ERK network 
 with the same phosphatase for MEK and ERK. (C)The ERK network with a negative 
 feedback. }\label{fig:3networks}
\end{figure}

The three networks are summarized in Figure~\ref{fig:3networks}.

In general, the existence of (positive) steady states and the capacity for 
multistationarity of chemical reaction systems is difficult to establish. 
Even for mass-action systems, the large number of interacting species and the 
lack of knowledge of the reaction rate constants become major drawbacks. If, 
however, the steady state ideal of the system is a binomial ideal, it was shown 
in \cite{psc11} -and recently generalized in \cite{mfrcsd13}- that these questions 
can be answered easily. Such systems are said to have \emph{toric steady states}. 
For these networks there are necessary and sufficient conditions that allow to 
decide about multistationarity and they take the form of linear inequality systems 
(based on previous work by \cite{csgr05}). 

In this work we show that the three MAPK systems we study have  toric steady states, 
which allows us to exploit the results in \cite{psc11} for determining the existence 
of positive steady states and the capacity for multistationarity of each system. 
In fact, each one of the three systems has many choices of rate constants for which 
they show multistationarity. We present, in the corresponding section, a certain choice 
of reaction constants for which each system has two different stable steady states. 
We can moreover conclude that the negative feedback loop is not necessary for the presence 
of bistability and neither does it prevent the system from this characteristic.

A similar mathematical analysis to signaling networks has been done in previous works. 
In \cite{cf12}, the authors present necessary and sufficient 
conditions for multistationarity for mass-action networks with certain structural 
properties and they also apply their results on some simple ERK cascade networks. 
The possibility of these networks having toric steady states is not taken into account, 
while we do consider this characteristic of the systems, thus simplifying the 
way to prove multistationarity.
In \citet{hfc13}, the authors describe a sign condition that is necessary and sufficient 
for multistationarity in $n$-site sequential, distributive phosphorylation.

Multistationarity in signaling pathways has also been studied in \citep{fw12,feliu12}. 
In the former, the authors study small motifs that repeatedly occur in these pathways. 
They include examples of a cascade with monostationarity and another with multistationarity, 
and although different tools are used for this result, it is possible to check that both systems 
have toric steady states. In \cite{feliu12}, the focus is on a signaling cascade with $n$ 
layers and one cycle of post-translational modification at each layer, such that the 
modified protein of one layer acts as modifier in the next layer, which is shown to 
have one steady state for fixed total amounts of substrates and enzymes. The analysis
is based on variable elimination, but it could be shown that 
these types of cascades also have toric steady states.

In our work we show that biologically relevant networks have toric steady states, 
which simplifies the analysis for multistationarity in the sense that it translates 
the question of finding two different (nonnegative) solutions of a system of 
polynomial equations into solving systems of linear inequalities.

We give in Section~\ref{sec:methods} the theoretical background needed 
to study the capacity for multistationarity of the ERK cascades via 
toric steady states. The main theorem is adapted from \cite{psc11}.
We then apply in Section~\ref{sec:results} our results to three specific 
ERK cascades presented in Figure~\ref{fig:3networks}: the standard ERK 
cascade; the ERK cascade with the same phosphatase for the MEK and ERK 
layers; and the ERK cascade with the same phosphatase for the MEK and 
ERK layers,  and a negative feedback loop. We show that we can find reasonable 
reaction constants and concentrations that allow to identify two different stable
steady states by modeling with ordinary differential equations. 
An appendix contains the details of the computations.

\FloatBarrier

\section{METHODS}\label{sec:methods}

We start this section with a brief presentation of the corresponding notation 
and we finish by revisiting the theorem obtained in \cite{psc11} which we 
will use to prove the capacity for multistationarity of the general MAPK's 
signaling networks with and without feedback. We model our networks under 
mass-action kinetics.

We introduce the notation with an example: the network for the smallest cascade.

\begin{example}\label{ex:simple}

For the 2-layer cascade of one cycle of post-translational modification at each layer 
we have the following network:

\begin{center}
\begin{tikzpicture}[node distance=1cm]
 \node[] at (-1,0) (dummy) {};
 \node[left=of dummy] (s0) {$S_0$};
  \node[right=of s0] (s1) {$S_1$}
   edge[->, bend left=45] node[below] {$F$} (s0)
   edge[<-, bend right=45] node[above] {$E$} (s0);
 \node[] at (1.1,-1.5) (dummy2) {};
 \node[left=of dummy2] (p0) {$P_0$};
 \node[right=of p0] (p1) {$P_1$}
   edge[->, bend left=45] node[below] {$F$} (p0)
   edge[<-, bend right=45] node[above] {$S_1$} (p0); 
\end{tikzpicture}
\end{center}

\medskip

where we consider the reactions:

\medskip

\begin{tabular}{llll}
 $S_0 + E \underset{k_2}{\overset{k_1}{\rightleftarrows}} ES_0 \overset{k_3}{\rightarrow} S_1 + E$ & & & \\
  & & \multicolumn{2}{l}{$S_1 + F \underset{k_5}{\overset{k_4}{\rightleftarrows}} FS_1 \overset{k_6}{\rightarrow} S_0 + F$}\\
 \multicolumn{2}{l}{$P_0 + S_1 \underset{k_8}{\overset{k_7}{\rightleftarrows}} S_1P_0 \overset{k_9}{\rightarrow} P_1 + S_1$}
  & & \\ \\
  & & \multicolumn{2}{l}{$P_1 + F \underset{k_{11}}{\overset{k_{10}}{\rightleftarrows}} FP_1 \overset{k_{12}}{\rightarrow} P_0 + F$}\\ \\
\end{tabular}

\end{example}

The network in Example~\ref{ex:simple} consists of ten \emph{species} $S_0$, $S_1$, $P_0$, $P_1$, $E$, $F$, 
$ES_0$, $S_1P_0$, $FS_1$ and $FP_1$ and twelve \emph{complexes}: $S_0+E$, $ES_0$, $S_1+E$, $S_1+F$, $FS_1$, 
$S_0+F$, $P_0+S_1$, $S_1P_0$, $P_1+S_1$, $P_1+F$, $FP_1$ and $P_0+F$. These complexes are connected by $12$ 
\emph{reactions}, where each reaction is associated with a \emph{rate constant} $k_i$. In the ordering chosen 
here, the first reaction would be $S_0+E\overset{k_1}{\rightarrow}ES_0$ with rate constant $k_1$. 
In this reaction, the complex $S_0+E$ reacts to the complex $ES_0$, hence $S_0 +E$ is called 
\emph{educt complex} and $ES_0$ \emph{product complex}.

We denote with $[\cdot]$ the concentration of a species and then correspond to each 
concentration a variable $x_i$. For example, we can consider:

\begin{small}
 $x_1 \leftrightarrow [S_0]$,  $x_2 \leftrightarrow [S_1]$, $x_3 \leftrightarrow [P_0]$,
 $x_4 \leftrightarrow [P_1]$,  $x_5 \leftrightarrow [E]$, $x_6 \leftrightarrow [F]$, 
 
 $x_7 \leftrightarrow [ES_0]$,  $x_8 \leftrightarrow [S_1P_0]$, $x_9 \leftrightarrow [FS_1]$, 
 and $x_{10} \leftrightarrow [FP_1]$. 
\end{small}

We associate to each species the corresponding canonical vector of 
$\mathbb{R}^{10}$ ($e_1$ to $S_0$, $e_2$ to $S_1$, $\dots$). Then 
every complex can be represented by the sum of its constituent 
species (use $y_i$ to denote \emph{complex vectors}): $y_1=e_1+e_5$ 
for $S_0+E$, and so on. 

Let us call $s$ the number of species, $m$ the number of complexes, and $r$ the number of reactions. Which, 
for the running example, would be $s=10$, $m=12$, and $r=12$.

\bigskip

Regarding the equations that describe the dynamics of the biochemical network,  under mass-action kinetics, 
reactions contribute production and consumption terms consisting of monomials like $k_1x_1x_5$ to the rates 
of formation of the species in the network. This results in a system of ordinary differential equations (ODEs), 
$dx/dt = f(x;k)$, in which each component rate function $f_i(x;k)$ is a polynomial in the state variables 
$x_1, x_2, \cdots, x_s \in \R$ and $k_1, \cdots, k_r \in \Rpl$ are positive rate constants.

The steady states of such ODEs are then zeros of a set of \emph{polynomial} equations, 
$f_1(x,k) = 0, \cdots, f_s(x,k) = 0$. Computational algebra and algebraic geometry provide 
powerful tools for studying these solutions \citep{clos}, and these tools have recently been 
used to gain new biological insights, for instance in \citep{rg08,mg09,mg07,cdss08,dcg10}. 
The rate constants can now be treated as symbolic parameters, whose numerical values do not 
need to be known in advance. The capability to rise above the parameter problem allows 
more general results to be obtained than can be expected from numerical simulation 
\citep{mg07,kpmddg12}.

The {\em steady state ideal} is defined as the set 
    \begin{align*}
      J~&=~\langle f_1,f_2, \dots, f_s \rangle \\
      &= \left\{ \sum_{i=1}^s g_i(x) f_i(x) \quad | \quad  g_i(x) \in
        \mathbb{R}[x_1,\dots, x_s] ~{\rm for }~  1 \leq i \leq s \right\}.
    \end{align*}
We say that the polynomial dynamical system has {\em toric steady states} if $J$ is a binomial 
ideal (i.e. the ideal $J$ can be generated by binomials) and it admits nonnegative zeros.

\bigskip

We will now introduce some matrices and subspaces that will be useful for studying multistationarity.

\medskip

The {\em stoichiometric subspace} is the vector subspace spanned by the {\em reaction vectors} 
$y_k-y_j$ (where there is a reaction from complex $y_j$ to complex $y_k$), and we will denote 
this space by $\mathcal{S}$.
We define the \emph{stoichiometric matrix}, $N$, in the following way: if the educt complex of the 
$i$-th reaction is $y_j$, and the product complex is $y_k$, then the $i$-th column of $N$ is the 
reaction vector $y_k-y_j$. Hence, $N$ is an $s\times r$ matrix.
Notice that $\mathcal{S}$ is exactly $\textrm{Columnspan}(N)$ 
(i.e. the subspace generated by the columns of $N$).
For the network in Example~\ref{ex:simple}, we obtain:

\begin{align*}
N=\left(\begin{array}{rrrrrrrrrrrr}
-1&1 &0 &0 &0 &1 &0 &0 &0 &0 &0 &0\\
0 &0 &1 &-1&1 &0 &-1&1 &1 &0 &0 &0\\
0 &0 &0 &0 &0 &0 &-1&1 &0 &0 &0 &1\\
0 &0 &0 &0 &0 &0 &0 &0 &1 &-1&1 &0\\
-1&1 &1 &0 &0 &0 &0 &0 &0 &0 &0 &0\\
0 &0 &0 &-1&1 &1 &0 &0 &0 &-1&1 &1\\
1 &-1&-1&0 &0 &0 &0 &0 &0 &0 &0 &0\\
0 &0 &0 &0 &0 &0 &1 &-1&-1&0 &0 &0\\
0 &0 &0 &1 &-1&-1&0 &0 &0 &0 &0 &0\\
0 &0 &0 &0 &0 &0 &0 &0 &0 &1 &-1&-1
\end{array}\right)
\end{align*}
If $\tilde y_i$ is the vector of the educt complex of the $i$-th reaction, we can define the vector 
of educt complex monomials
\begin{equation*}
  \phi(x) ~:=~ \left(x^{\tilde y_1}, ~x^{\tilde y_2},~ \ldots, ~x^{\tilde y_r}\right)^t~.
\end{equation*}
In our example, this vector would be:
$$\phi(x)=(x_1x_5,x_7,x_7,x_2x_6,x_9,x_9,x_2x_3,x_8,x_8,x_4x_6,x_{10},x_{10})^t.$$
We also define $k \in \mathbb{R}^r_{>0}$ to be the vector of reaction rate constants: 
$k_i$ is the rate constant of the $i$-th reaction.
A chemical reaction system can then be expressed as:
\begin{equation*}
  \dot x ~=~ N\, \diag(k)\, \phi(x)\ .
\end{equation*}
The vector $\dot x$  lies in $\mathcal{S}$ for all time $t$. In fact, a trajectory $x(t)$ beginning 
at a positive vector $x(0)=x^0 \in \R^s_{>0}$ remains in the {\em stoichiometric compatibility class} 
$(x^0+\mathcal{S}) \cap \mathbb{R}^s_{\geq 0}$ for all positive time. The equations of $x^0+\mathcal{S}$
give rise to the \emph{conservation relations} of the system.

In our example, the conservation relations are:
\begin{align}
 \nonumber x_1+x_2+x_7+x_8+x_9&=C_1\\\label{eq:cons_rel}
  x_3+x_4+x_8+x_{10}&=C_2\\ 
 \nonumber x_5+x_7&=C_3\\
 \nonumber x_6+x_9+x_{10}&=C_4
\end{align}
These conservation relations in~\eqref{eq:cons_rel} can be translated as the conservation of the total amounts 
of the first-layer substrate, $S$, the second-layer substrate, $P$, and the enzymes $E$ and $F$, respectively.

A chemical reaction system exhibits {\em multistationarity} if there exists a stoichiometric
compatibility class with two or more steady states in its relative interior. A system
may admit multistationarity for all, some, or no choices of positive rate constants $k_i$; 
if such rate constants exist, then we say that the network {\em has the capacity for multistationarity}.

We now recognize that the set $\ker(N)\cap\mathbb{R}_{>0}^r$, if nonempty, is the relative interior 
of the pointed polyhedral cone $\ker(N)\cap\mathbb{R}_{\geq 0}^r$. To utilize this cone, we collect 
a finite set of generators (also called ``extreme rays'') of the cone 
$\ker(N)\cap\mathbb{R}_{\geq 0}^r$ as columns of a non-negative matrix $M$.

For network in Example~\ref{ex:simple}, a possible matrix $M$ is:
$$M=\left(
\begin{array}{rrrrrr}
 1&0&0&0&1&0\\
 1&0&0&0&0&0\\
 0&0&0&0&1&0\\
 0&1&0&0&1&0\\
 0&1&0&0&0&0\\
 0&0&0&0&1&0\\
 0&0&1&0&0&1\\
 0&0&1&0&0&0\\
 0&0&0&0&0&1\\
 0&0&0&1&0&1\\
 0&0&0&1&0&0\\
 0&0&0&0&0&1
\end{array}\right)$$
If the steady state ideal $J$ is generated by the binomials $b_ix^{\hat y_j}-b_jx^{\hat y_i}$, 
let $A \in \mathbb{Z}^{w \times s}$ be a matrix of maximal rank such that $\ker(A)$ equals 
the span of all the differences $\hat y_j-\hat y_i$.
For the mass-action system arising from the network in Example~\ref{ex:simple}, 
the ideal $J$ can be generated by the binomials 
$$\begin{array}{lcl}
   k_1x_1x_5-(k_2+k_3)x_7 & & k_{10}x_4x_6-(k_{11}+k_{12})x_{10}\\
   k_4x_2x_6-(k_5+k_6)x_9 & & k_3x_7-k_6x_9\\
   k_7x_2x_3-(k_8+k_9)x_8 & & k_9x_8-k_{12}x_{10}
  \end{array}$$
Then, a possible matrix $A$ is 
$$A=\left(\begin{array}{cccccccccc}
           1 & 1 & 0 & 1 & 0 & 0 & 1 & 1 & 1 & 1\\
           1 & 0 & 1 & 0 & 0 & 1 & 1 & 1 & 1 & 1\\
           0 & 0 & 1 & 1 & 0 & 0 & 0 & 1 & 0 & 1\\
           0 & 0 & 1 & 0 & 1 & 1 & 1 & 1 & 1 & 1
          \end{array}\right)$$
We define the sign of a vector $v\in\R^s$ as a vector $sign(v) \in \{-,0,+\}^s$ whose $i$-th coordinate is 
the sign of the $i$-th entry of $v$.

The following theorem from \cite{psc11} is the one that we will use to study the multistationarity of 
MAPK's networks with and without feedback.

\bigskip

\begin{theorem}\label{thm:teo}
 Given matrices $A$ and $N$ as above, and nonzero vectors $\alpha\in\textrm{Rowspan}(A)$
    and $\sigma\in\textrm{Columnspan}(N)$ with
    \begin{equation}
      \label{eq:condi}
      \sign(\alpha)~=~\sign(\sigma) ~, 
    \end{equation}
  then two steady states $x^1$ and $x^2$ and a reaction rate constant vector $k$ 
    that witness multistationarity arise in the following way:
    \begin{align}
      \label{eq:def_x1}
      \left(x^1_i\right)_{i=1,\, \ldots,\, s} ~&=~ 
      \begin{cases}
        \frac{\sigma_{i}}{e^{\alpha_{i}}-1}\text{, if $\alpha_{i} \neq
          0$} \\ 
        \bar{x}_i>0\text{, if $\alpha_{i} = 0$~,}
      \end{cases}
      \intertext{where $\bar{x}_i$ denotes an arbitrary positive number, 
        and}
      \label{eq:def_x2}
      x^2 ~&=~ \diag(e^\alpha)\, x^1 \\
      \label{eq:def_k}
      k ~&=~ \diag(\phi(x^1))^{-1}\, M \, \lambda~ ,
    \end{align}
    for any non-negative vector $\lambda\in\mathbb{R}_{\geq 0}^p$ for which 
    $ M \, \lambda \in \mathbb{R}_{>0}^r$.  Conversely, any witness to multistationarity
    (given by some $x^1$, $x^2\in\mathbb{R}_{>0}^s$, and $k\in\mathbb{R}_{>0}^r$)  
    arises from equations~ \eqref{eq:condi}, 
    \eqref{eq:def_x1},
    \eqref{eq:def_x2}, and
    \eqref{eq:def_k} for some 
    vectors $\alpha\in\textrm{Rowspan}(A)$
    and $\sigma\in\textrm{Columnspan}(N)$ that have the same sign.
\end{theorem}

\section{RESULTS}\label{sec:results}

We prove in this section the capacity for multistationarity of three 
networks that are frequently used to represent the principal kinase 
transduction pathways in eukaryotic cells, which are the MAPK cascades. 
The first network is the most frequent in the literature \citep{kholo00,ferrell96}. 
The second one differs from the first one in the phosphatases, which 
we assume to be equal for the last two layers of the cascades \citep{fujioka}. 
The third network includes a negative feedback between pRAF and ppERK in which 
the latter acts as a kinase for the former, producing a new phosphorylated and 
inactive form Z \citep{al01,dough05,fritsche}. 
The three networks are summarized in Figure~\ref{fig:3networks}.

We determine that the ERK cascades present toric steady states, and this helps us to 
prove the capacity for multistationarity of each system. We then  determine reaction 
constants and concentrations that witness multistability.
Namely, we analyze the ODEs that arise under mass-action for each network, and we find 
that the corresponding steady state ideals are binomial. We show, in different appendices, 
an order for the species of each network, the conservation relations, and binomials that 
generate the mentioned ideal. We also present a matrix $A$ as in Section~\ref{sec:methods} 
for studying multistationarity, and we include the corresponding matrices $N$ and $M$, 
and vector $\phi(x)$. By solving three different systems of sign equalities, we find vectors 
$\alpha \in \textrm{Rowspan}(A)$ and $\sigma \in \mathcal{S}$ with 
$\textrm{sg}(\alpha_i)=\textrm{sg}(\sigma_i)$ (for each system) as required by 
Theorem~\ref{thm:teo} for proving the capacity for multistationarity. With the 
aid of these vectors, we can build two different steady states and a vector of 
reaction constants which, according to Theorem~\ref{thm:teo}, witness to 
multistationarity in each case in the corresponding stoichiometric compatibility 
class defined by the constants (i.e. total amounts). It 
can be checked that the steady states we find are stable.

In the following subsections we treat each network separately.
Numerical computations and simulations in this article were performed with~\citeauthor{matlab}, 
while computations regarding ideals and subspaces were done with~\citeauthor{singular}.

\subsection{The network without feedback and three phosphatases} \label{subsec:red1}

We start by studying the network for the signaling pathway of 
ERK without feedback (see Figure~\ref{fig:3networks}(A);\citet{kholo00,ferrell96}). This 
network entails $s=22$ species, $m=26$ complexes and 
$r=30$ reactions which are as follows:

\noindent \begin{tabular}{l}
{\scriptsize RAF + RAS} 
$\underset{k_{2}}{\overset{k_1}{\rightleftarrows}}$ 
{\scriptsize RAS-RAF}
$\overset{k_3}{\rightarrow}$
{\scriptsize pRAF + RAS}\\
 
{\scriptsize pRAF + RAFPH} 
$\underset{k_{5}}{\overset{k_4}{\rightleftarrows}}$
{\scriptsize RAF-RAFPH} 
$\overset{k_6}{\rightarrow}$
{\scriptsize RAF + RAFPH} \\
 
{\scriptsize MEK + pRAF} 
$\underset{k_{8}}{\overset{k_7}{\rightleftarrows}}$
{\scriptsize MEK-pRAF} 
$\overset{k_9}{\rightarrow}$
{\scriptsize pMEK + pRAF}
$\underset{k_{11}}{\overset{k_{10}}{\rightleftarrows}}$ 
{\scriptsize pMEK-pRAF}
$\overset{k_{12}}{\rightarrow}$ 
{\scriptsize ppMEK + pRAF}\\
 
{\scriptsize ppMEK+MEKPH}
$\underset{k_{14}}{\overset{k_{13}}{\rightleftarrows}}$ 
{\scriptsize ppMEK-MEKPH} 
$\overset{k_{15}}{\rightarrow}$
{\scriptsize pMEK+MEKPH}
$\underset{k_{17}}{\overset{k_{16}}{\rightleftarrows}}$ 
{\scriptsize pMEK-MEKPH} 
$\overset{k_{18}}{\rightarrow}$ 
{\scriptsize MEK+MEKPH}\\
 
{\scriptsize ERK+ppMEK} 
$\underset{k_{20}}{\overset{k_{19}}{\rightleftarrows}}$ 
{\scriptsize ERK-ppMEK} 
$\overset{k_{21}}{\rightarrow}$
{\scriptsize pERK+ppMEK} 
$\underset{k_{23}}{\overset{k_{22}}{\rightleftarrows}}$ 
{\scriptsize pERK-ppMEK} 
$\overset{k_{24}}{\rightarrow}$
{\scriptsize ppERK+ppMEK}\\
 
{\scriptsize ppERK+ERKPH} 
$\underset{k_{26}}{\overset{k_{25}}{\rightleftarrows}} $
{\scriptsize ppERK-ERKPH }
$\overset{k_{27}}{\rightarrow} $
{\scriptsize pERK+ERKPH}
$\underset{k_{29}}{\overset{k_{28}}{\rightleftarrows}}$
{\scriptsize pERK-ERKPH}
$\overset{k_{30}}{\rightarrow}$
{\scriptsize ERK+ERKPH}
\end{tabular}

\medskip
 
\begin{figure}[ht]

\begin{tabular}{cc}
 \small{(a)}\includegraphics[scale=0.32,trim=1.7cm 6.8cm 2cm 6.7cm, clip=true]{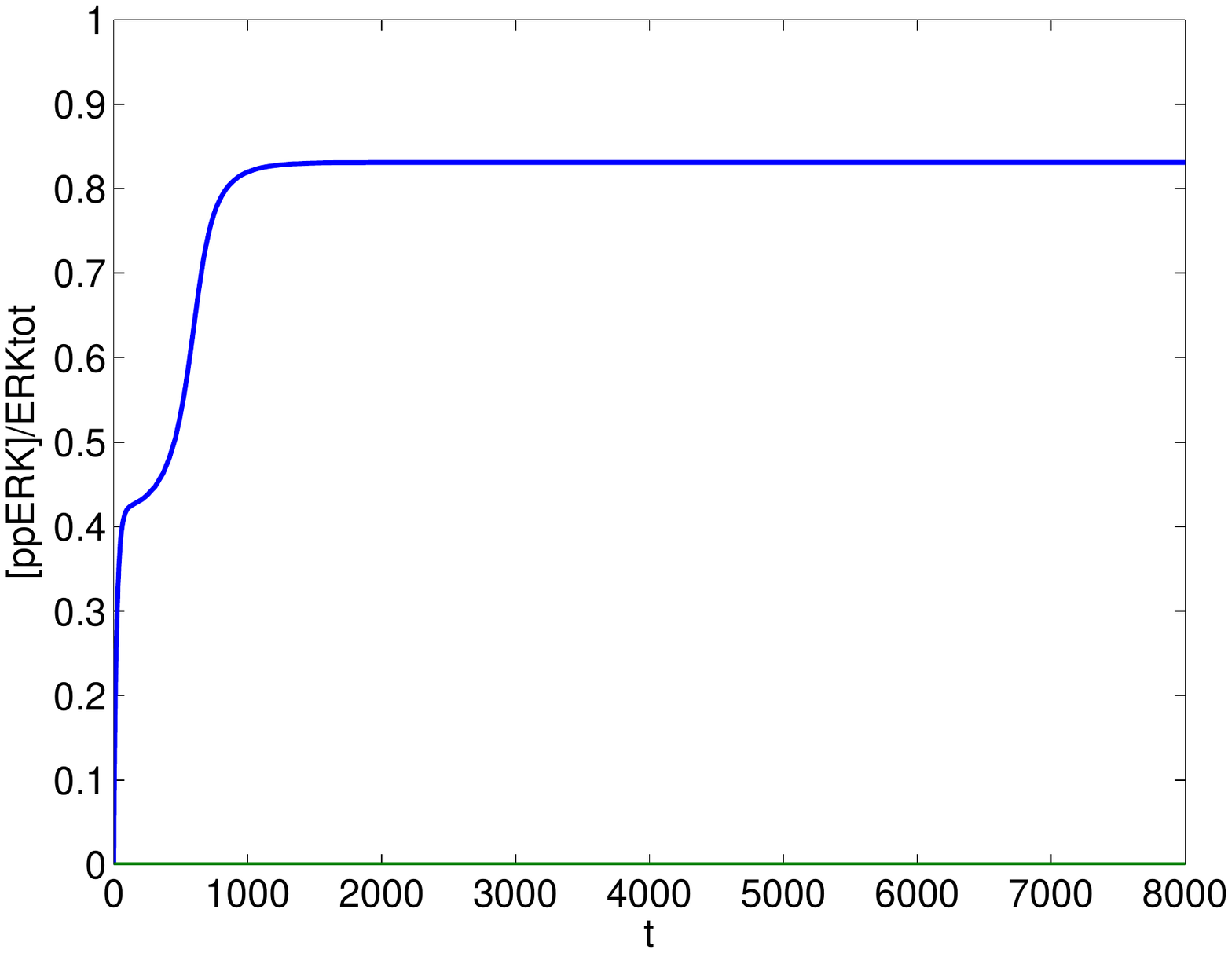} &
 \small{(b)}\includegraphics[scale=0.32,trim=1.7cm 6.8cm 2cm 7cm, clip=true]{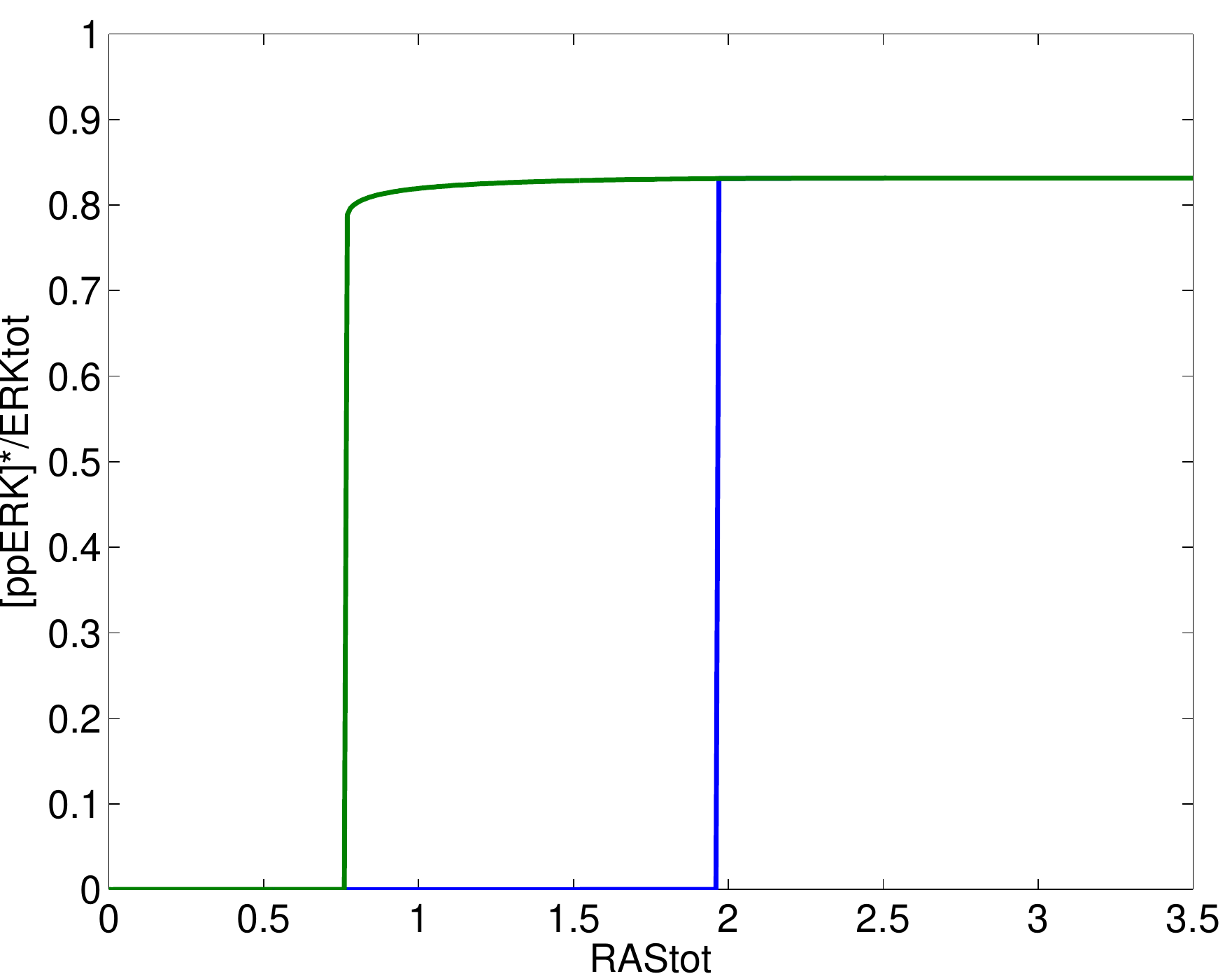} 
\end{tabular}

\caption{ (a) The normalized trajectory of ppERK vs. time for two different initial values 
in the same stoichiometric compatibility class. The nonzero entries of the initial value for the green curve are  
[RAF]=12.8629, [MEK]=11.9697, [ERK]=23.3465, [RAS]=2, [RAFPH]=2, [MEKPH]=7.3058 and [ERKPH]=3.2013. 
The nonzero entries of the initial value for the blue curve are the same except for 
[MEK]=4.5697 and [ppMEK]=7.4.
(b) Dose-response curve for the network without feedback. 
 The horizontal axis represents the total amount of the dose ([RAS]+[RAS-RAF]),
 and the vertical axis stands for the normalized equilibrium values of [ppERK]. For 
 each value of RAS, the corresponding equilibria belong to the same 
 stoichiometric compatibility class.}
\label{fig:n1}
\end{figure}

We can prove that the corresponding mass-action system is 
capable of reaching two significantly different (stable) steady states 
in the same stoichiometric compatibility class. We
refer the reader to Appendix A for the corresponding computations.
Figure~\ref{fig:n1} pictures this feature of the system.

\FloatBarrier

\subsection{The network without feedback and two phosphatases}  \label{subsec:red2}

We now study the network for the signaling pathway of 
ERK without feedback and the same phosphatase for both, MEK and ERK 
(see Figure~\ref{fig:3networks}(B); \citet{fujioka}). This network entails $s=21$ species, 
$m=26$ complexes and $r=30$ reactions which are:

\noindent \begin{tabular}{l}
{\scriptsize RAF + RAS} 
$\underset{k_2}{\overset{k_1}{\rightleftarrows}}$ 
{\scriptsize RAS-RAF}
$\overset{k_3}{\rightarrow}$
{\scriptsize pRAF + RAS}\\
 
{\scriptsize pRAF + RAFPH} 
$\underset{k_5}{\overset{k_4}{\rightleftarrows}}$
{\scriptsize RAF-RAFPH} 
$\overset{k_6}{\rightarrow}$
{\scriptsize RAF + RAFPH} \\
 
{\scriptsize MEK + pRAF} 
$\underset{k_8}{\overset{k_7}{\rightleftarrows}}$
{\scriptsize MEK-pRAF} 
$\overset{k_9}{\rightarrow}$
{\scriptsize pMEK + pRAF}
$\underset{k_{11}}{\overset{k_{10}}{\rightleftarrows}}$ 
{\scriptsize pMEK-pRAF}
$\overset{k_{12}}{\rightarrow}$ 
{\scriptsize ppMEK + pRAF}\\
 
{\scriptsize ppMEK+PH}
$\underset{k_{14}}{\overset{k_{13}}{\rightleftarrows}}$ 
{\scriptsize ppMEK-PH} 
$\overset{k_{15}}{\rightarrow}$
{\scriptsize pMEK+PH}
$\underset{k_{17}}{\overset{k_{16}}{\rightleftarrows}}$ 
{\scriptsize pMEK-PH} 
$\overset{k_{18}}{\rightarrow}$ 
{\scriptsize MEK+PH}\\
 
{\scriptsize ERK+ppMEK} 
$\underset{k_{20}}{\overset{k_{19}}{\rightleftarrows}}$ 
{\scriptsize ERK-ppMEK} 
$\overset{k_{21}}{\rightarrow}$
{\scriptsize pERK+ppMEK} 
$\underset{k_{23}}{\overset{k_{22}}{\rightleftarrows}}$ 
{\scriptsize pERK-ppMEK} 
$\overset{k_{24}}{\rightarrow}$
{\scriptsize ppERK+ppMEK}\\
 
{\scriptsize ppERK+PH} 
$\underset{k_{26}}{\overset{k_{25}}{\rightleftarrows}} $
{\scriptsize ppERK-PH }
$\overset{k_{27}}{\rightarrow} $
{\scriptsize pERK+PH}
$\underset{k_{29}}{\overset{k_{28}}{\rightleftarrows}}$
{\scriptsize pERK-PH}
$\overset{k_{30}}{\rightarrow}$
{\scriptsize ERK+PH}
\end{tabular}
    
\medskip    

\begin{figure}[ht]

\begin{tabular}{cc}
 \small{(a)}\includegraphics[scale=0.32,trim=1.7cm 6.8cm 2cm 6.7cm, clip=true]{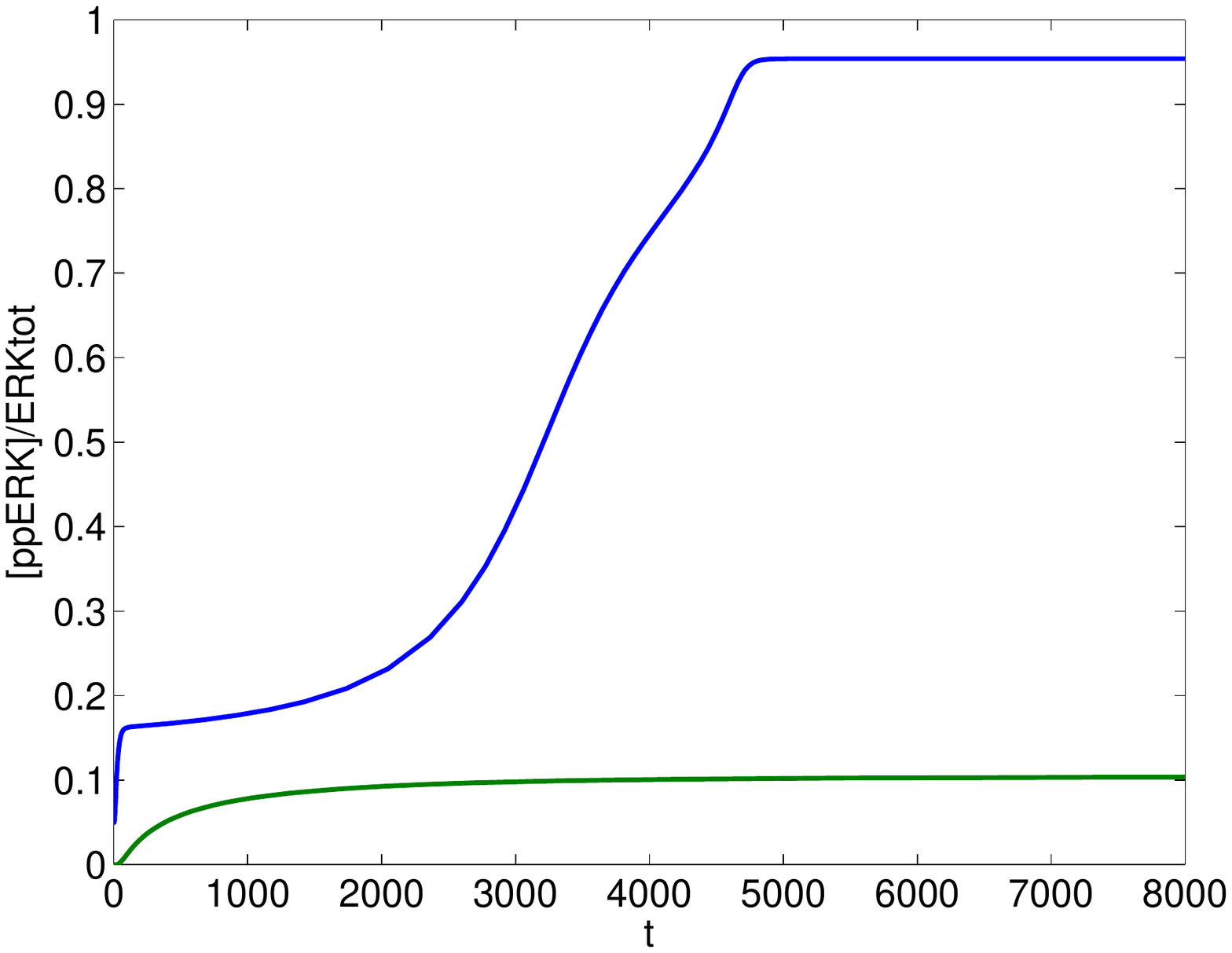} &
 \small{(b)}\includegraphics[scale=0.32,trim=1.7cm 6.8cm 2cm 7cm, clip=true]{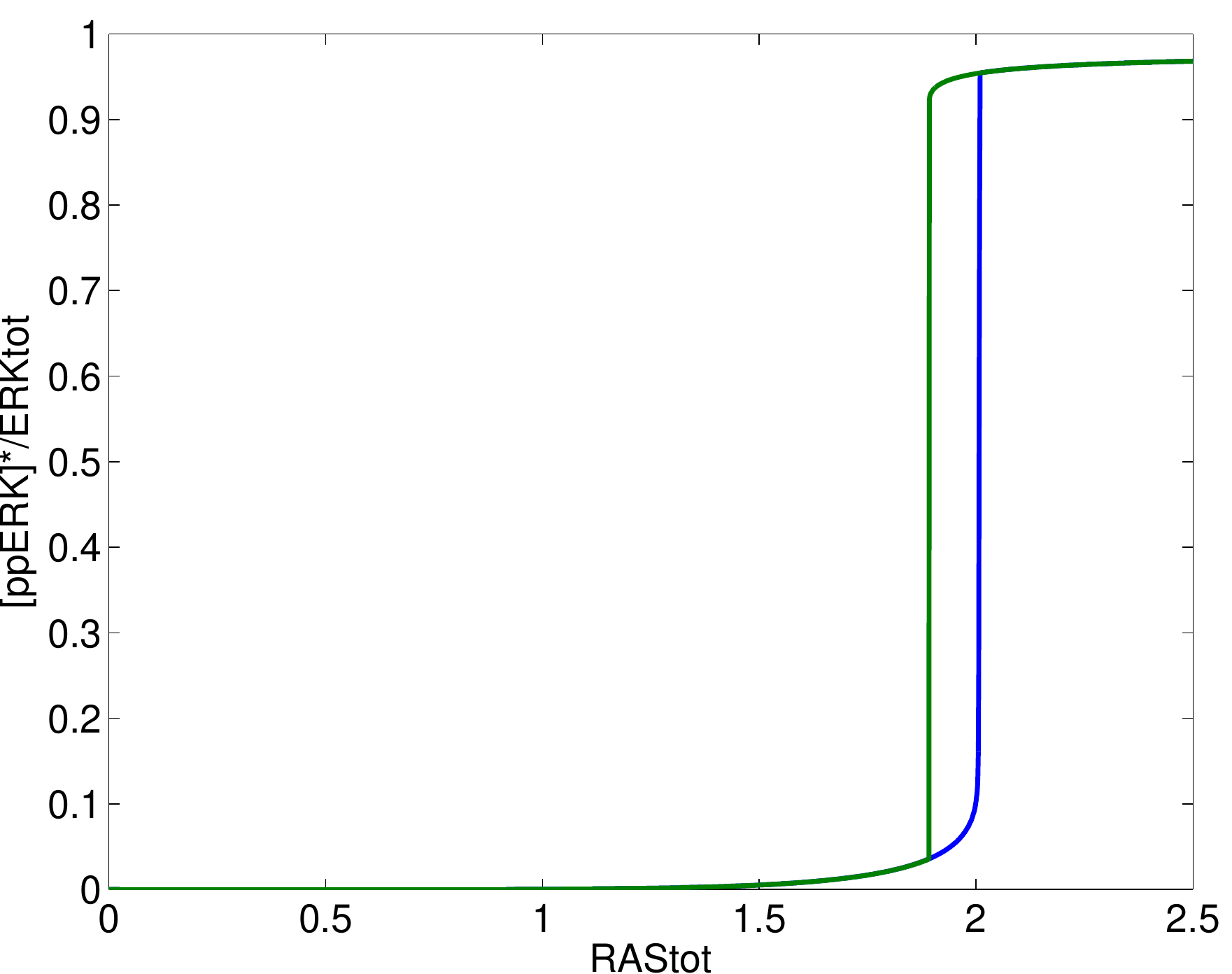}
\end{tabular}

\caption{ For the network without feedback and the same phosphatase 
for MEK and ERK. (a) The normalized trajectory of ppERK vs. time for two different initial values 
in the same stoichiometric compatibility class. The nonzero entries of the initial value for the green curve are  
[RAF]=4.1738, [MEK]=1.9063, [ERK]=3.7737, [RAS]=2, [RAFPH]=2 and [PH]=1.3995. 
The nonzero entries of the initial value for the blue curve are the same except for 
[MEK]=0.0063, [ppMEK]=1.9, [ERK]=3.5237 and [ppERK]=0.25.
(b) Dose-response curve for the network without feedback. 
 The horizontal axis represents the total amount of the dose ([RAS]+[RAS-RAF]),
 and the vertical axis stands for the normalized equilibrium values of [ppERK]. For 
 each value of RAS, the corresponding equilibria belong to the same 
 stoichiometric compatibility class.}
\label{fig:n2}
\end{figure}

We depict in Figure~\ref{fig:n2} the hysteresis and 
bistability this network presents. All the necessary information for this network 
is presented in Appendix~B.

\FloatBarrier

\subsection{The network with feedback}  \label{subsec:red3}

We now study a network for the signaling pathway of 
ERK with a negative feedback  between pRAF and ppERK in which the latter 
acts as a kinase for the former, producing a new phosphorylated and inactive form Z
(see Figure~\ref{fig:3networks}(C);\citet{al01,dough05,fritsche} ). This network consists of $s=25$ species, $m=32$ complexes and 
$r=36$ reactions.

\noindent \begin{tabular}{l}
{\scriptsize RAF + RAS} 
$\underset{k_2}{\overset{k_1}{\rightleftarrows}}$ 
{\scriptsize RAS-RAF}
$\overset{k_3}{\rightarrow}$
{\scriptsize pRAF + RAS}\\
 
{\scriptsize pRAF + RAFPH} 
$\underset{k_5}{\overset{k_4}{\rightleftarrows}}$
{\scriptsize RAF-RAFPH} 
$\overset{k_6}{\rightarrow}$
{\scriptsize RAF + RAFPH} \\
 
{\scriptsize MEK + pRAF} 
$\underset{k_8}{\overset{k_7}{\rightleftarrows}}$
{\scriptsize MEK-pRAF} 
$\overset{k_9}{\rightarrow}$
{\scriptsize pMEK + pRAF}
$\underset{k_{11}}{\overset{k_{10}}{\rightleftarrows}}$ 
{\scriptsize pMEK-pRAF}
$\overset{k_{12}}{\rightarrow}$ 
{\scriptsize ppMEK + pRAF}\\
 
{\scriptsize ppMEK+PH}
$\underset{k_{14}}{\overset{k_{13}}{\rightleftarrows}}$ 
{\scriptsize ppMEK-PH} 
$\overset{k_{15}}{\rightarrow}$
{\scriptsize pMEK+PH}
$\underset{k_{17}}{\overset{k_{16}}{\rightleftarrows}}$ 
{\scriptsize pMEK-PH} 
$\overset{k_{18}}{\rightarrow}$ 
{\scriptsize MEK+PH}\\
 
{\scriptsize ERK+ppMEK} 
$\underset{k_{20}}{\overset{k_{19}}{\rightleftarrows}}$ 
{\scriptsize ERK-ppMEK} 
$\overset{k_{21}}{\rightarrow}$
{\scriptsize pERK+ppMEK} 
$\underset{k_{23}}{\overset{k_{22}}{\rightleftarrows}}$ 
{\scriptsize pERK-ppMEK} 
$\overset{k_{24}}{\rightarrow}$
{\scriptsize ppERK+ppMEK}\\
 
{\scriptsize ppERK+PH} 
$\underset{k_{26}}{\overset{k_{25}}{\rightleftarrows}} $
{\scriptsize ppERK-PH }
$\overset{k_{27}}{\rightarrow} $
{\scriptsize pERK+PH}
$\underset{k_{29}}{\overset{k_{28}}{\rightleftarrows}}$
{\scriptsize pERK-PH}
$\overset{k_{30}}{\rightarrow}$
{\scriptsize ERK+PH}\\

{\scriptsize pRAF + ppERK} 
$\underset{k_{32}}{\overset{k_{31}}{\rightleftarrows}}$ 
{\scriptsize pRAF-ppERK}
$\overset{k_{33}}{\rightarrow}$
{\scriptsize Z + ppERK}\\

{\scriptsize Z + PH2} 
$\underset{k_{35}}{\overset{k_{34}}{\rightleftarrows}}$ 
{\scriptsize Z-PH2}
$\overset{k_{36}}{\rightarrow}$
{\scriptsize pRAF + PH2}
\end{tabular}

\medskip

\begin{figure}[ht]

\begin{tabular}{cc}
 \small{(a)}\includegraphics[scale=0.32,trim=1.7cm 6.8cm 2cm 6.7cm, clip=true]{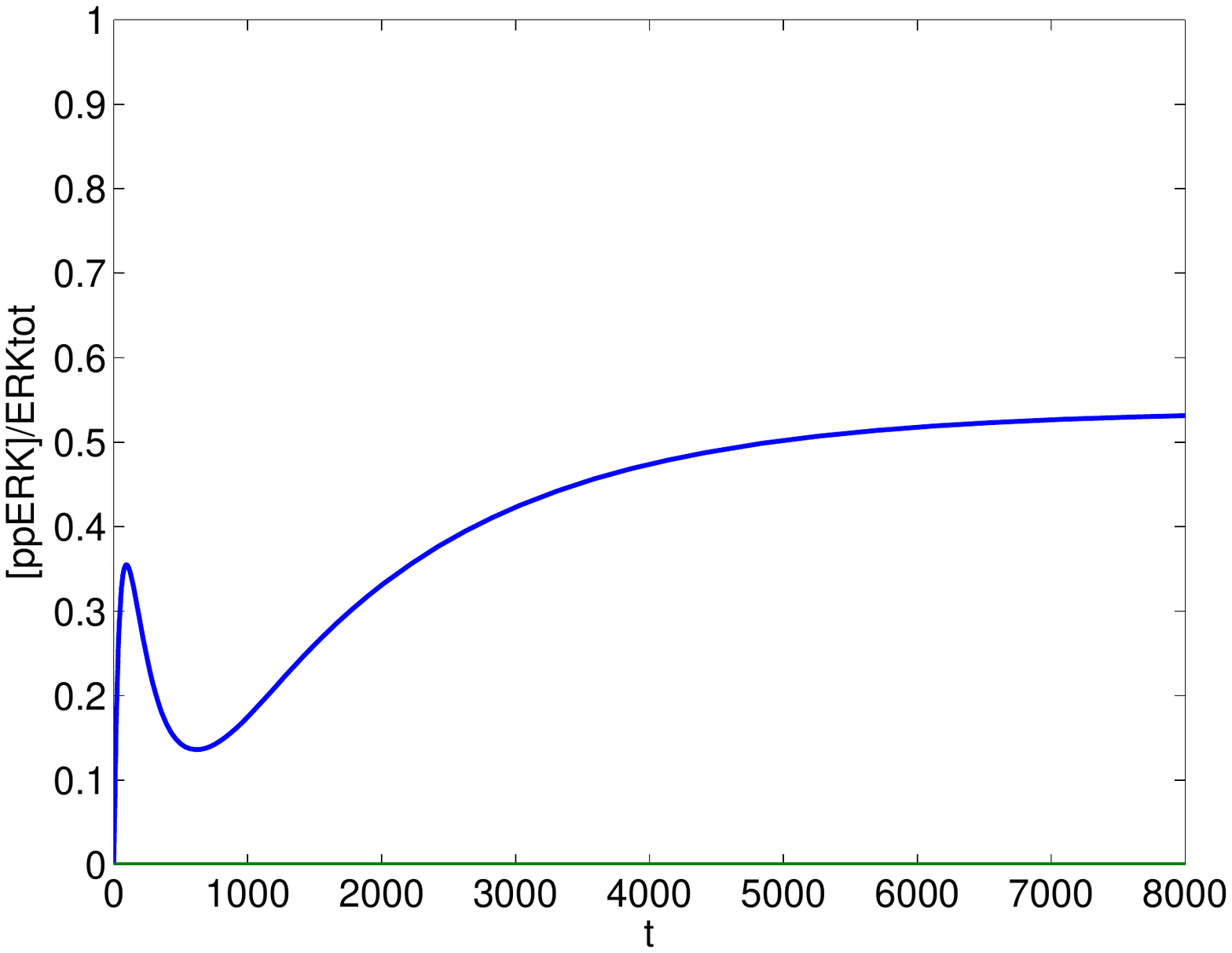} &
 \small{(b)}\includegraphics[scale=0.32,trim=1.7cm 6.8cm 2cm 7cm, clip=true]{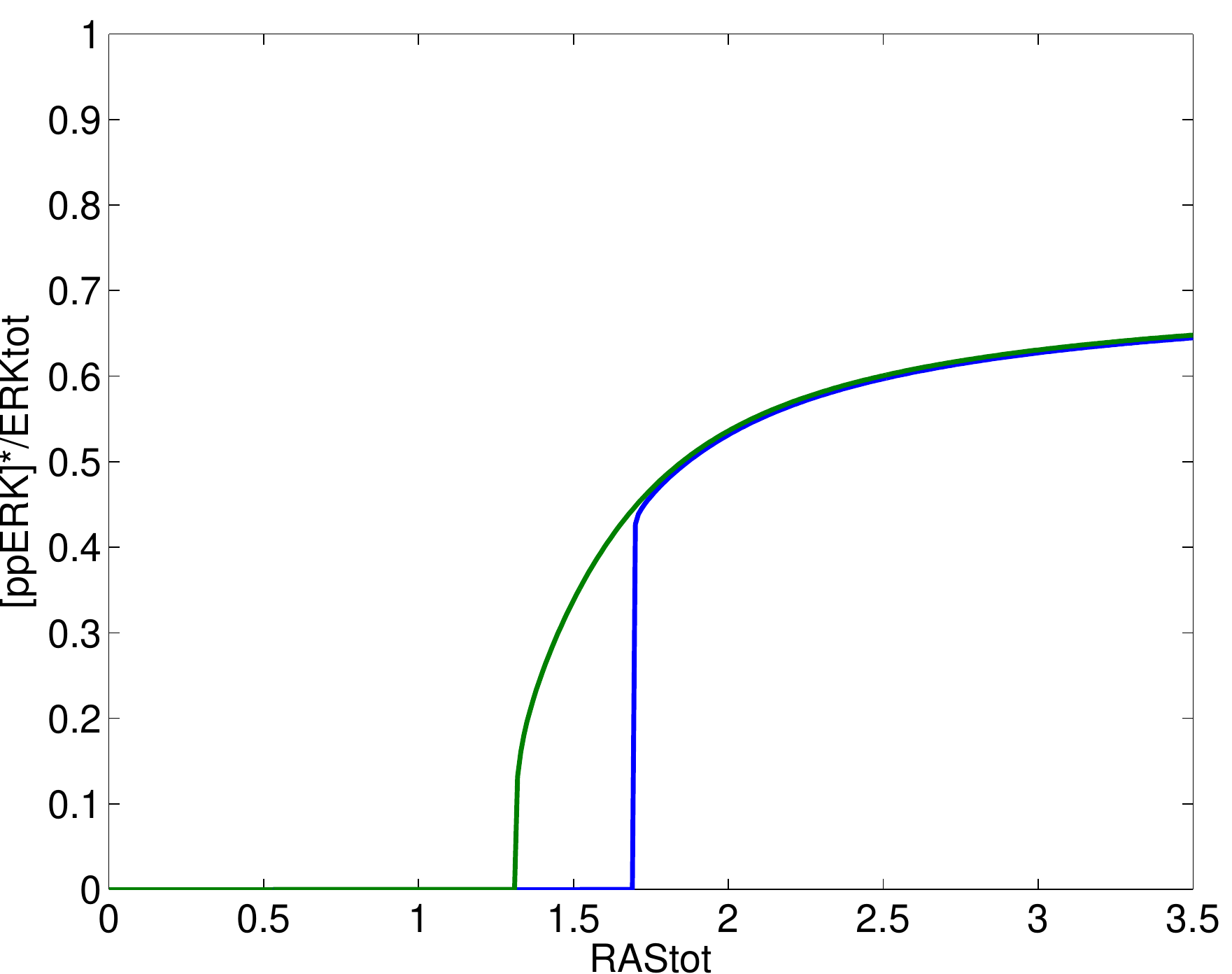}
\end{tabular}

\caption{ For the network with a negative feedback and the same phosphatase 
for MEK and ERK. (a) The normalized trajectory of ppERK vs. time for two different initial values 
in the same stoichiometric compatibility class. The nonzero entries of the initial value for the green curve are  
[RAF]=12.8629, [MEK]=11.9697, [ERK]=23.3465, [RAS]=2, [RAFPH]=2, [PH]=7.3058 and [PH2]=3.2013. 
The nonzero entries of the initial value for the blue curve are the same except for 
[MEK]=4.5697 and [ppMEK]=7.4.
(b) Dose-response curve for the network without feedback. 
 The horizontal axis represents the total amount of the dose ([RAS]+[RAS-RAF]),
 and the vertical axis stands for the normalized equilibrium values of [ppERK]. For 
 each value of RAS, the corresponding equilibria belong to the same 
 stoichiometric compatibility class.}
\label{fig:n3}
\end{figure}

We can prove that the corresponding mass-action system is 
capable of reaching two significantly different (stable) steady states 
in the same stoichiometric compatibility class. We
refer the reader to Appendix C for the corresponding computations.
Figure~\ref{fig:n3} pictures this feature of the system.

\section{DISCUSSION}

We have applied a useful algebraic tool for studying the capacity for 
multistationarity of an important signaling pathway as the MAPK cascade. 
We included in our analysis three frequent possible networks for describing 
the MAPK signaling mechanism, which happen to have, under mass-action 
kinetics, a binomial steady state ideal. This allowed us to translate the 
question of multistationarity to a system of sign equalities, and so  
we proved that ERK systems are able to show multistationarity for reasonable 
choices of rate constants. 

The application of computational biology and systems biology is yielding quantitative 
insight into cellular regulatory phenomena and a 
a large number of papers have appeared that 
estimate in-vivo protein concentrations and reaction constants of the MAPK signaling 
networks \citep{hornberg}. However, there is no agreement about these values
and differences of more than two orders of magnitude have appeared \citep{qiao07}.
In this sense, in our case, any nontrivial solution of the linear inequality system 
defined by~\eqref{eq:condi} gives two different steady sates and a set of rate constants 
for which the system has those steady states, and both the steady states and the constants 
are determined explicitly. Our results highlight that
the robustness of the topology also tolerates changes in protein concentrations and rate 
constants, allowing a similar overall behavior of the network. Concentrations may vary 
from one organism to another, and kinetic constants can be regulated by different mechanisms 
as for example the role of scaffolds in MAPK kinase cascades \citep{kolch05}.
We also reveal that the MAPK cascades are robust in the sense that neither the differences 
in phosphatases nor the presence or absence of feedback loops alter the capacity 
for multistability.

Finally, algebraic methods are proving to be powerful tools for answering questions 
from biochemical reaction network studies. In particular, they are very 
useful for addressing matters of steady state characterization 
\citep{kpmddg12,mfrcsd13}. The same analysis we performed in the present work 
could be applied to many other important biochemical networks as long as they 
present toric steady states. We are currently developing easier (graphical) 
methods for detecting this characteristic in enzymatic networks, and we plan 
to improve the computational methods for solving the system of sign equalities 
defined by Equation~\eqref{eq:condi}.

\section{ACKNOWLEDGEMENTS}
We thank C. Conradi for technical assistance in the first stages of this article
and E. Feliu for some useful suggestions.

MPM was partially supported by UBACYT  20020100100242, CONICET PIP 11220110100580 and ANPCyT 2008-0902, Argentina. 
This research was partially supported by UBACYT 20020110100061BA and ANPCyT 2010-2805.
AGT is a staff member of CONICET.

\appendix

\section{The ERK network without feedback.}

We present in this appendix the matrices, vectors, constants and corresponding 
(stable) steady states that prove the capacity for multistationarity 
for the system without feedback and three different phosphatases for 
each substrate, presented in Subsection~\ref{subsec:red1}. 

\medskip

The conservation relations of this system are:
\begin{footnotesize}
\begin{align*}
 \text{[RAF]+[pRAF]+[RAS-RAF]+[MEK-pRAF]+[pMEK-pRAF]+[pRAF-RAFPH]} &=C_1\\ 
\text{[MEK]+[pMEK]+[ppMEK]+[MEK-pRAF]+[pMEK-pRAF]+[ERK-ppMEK]+}&\\
\text{+[pERK-ppMEK]+[ppMEK-MEKPH]+[pMEK-MEKPH]}&=C_2\\
\text{[ERK]+[pERK]+[ppERK]+[ERK-ppMEK]+[pERK-ppMEK]+}&\\
\text{+[ppERK-ERKPH]+[pERK-ERKPH]}&=C_3\\
\text{[RAS]+[RAS-RAF]}&=C_4\\
\text{[RAFPH]+[RAF-RAFPH]}&=C_5\\
\text{[MEKPH]+[ppMEK-MEKPH]+[pMEK-MEKPH]}&=C_6\\
\text{[ERKPH]+[ppERK-ERKPH]+[pERK-ERKPH]}&=C_7
\end{align*}
\end{footnotesize}
Where $C_1, \dots, C_7$ usually stand, respectively, for {\small [RAF]$_{tot}$}, {\small [MEK]$_{tot}$},  
{\small [ERK]$_{tot}$}, {\small [RAS]$_{tot}$}, {\small [RAFPH]$_{tot}$}, {\small [MEKPH]$_{tot}$}, 
and {\small [ERKPH]$_{tot}$}, the total amounts of the corresponding species.

\medskip

Under mass-action kinetics, the steady state ideal for this network is binomial. In fact, if we consider
the following order of the species:

\begin{small}
 $x_1 \leftrightarrow [RAF]$,  $x_2 \leftrightarrow [pRAF]$,
 $x_3 \leftrightarrow [MEK]$, $x_4 \leftrightarrow [pMEK]$, $x_5 \leftrightarrow [ppMEK]$,
 
 $x_6 \leftrightarrow [ERK]$,  $x_7 \leftrightarrow [pERK]$,  $x_8 \leftrightarrow [ppERK]$,
 $x_9 \leftrightarrow [RAS]$, $x_{10} \leftrightarrow [RAFPH]$, 
 
 $x_{11} \leftrightarrow [MEKPH]$,  $x_{12} \leftrightarrow [ERKPH]$  
 $x_{13} \leftrightarrow [RAS-RAF]$,  
 
 $x_{14} \leftrightarrow [MEK-pRAF]$,  $x_{15} \leftrightarrow [pMEK-pRAF]$, $x_{16} \leftrightarrow [ERK-ppMEK]$,   
 
 $x_{17} \leftrightarrow [pERK-ppMEK]$,
 $x_{18} \leftrightarrow [RAF-RAFPH]$,  
 
 $x_{19} \leftrightarrow [ppMEK-MEKPH]$,  
 $x_{20} \leftrightarrow [pMEK-MEKPH]$,  
 
 $x_{21} \leftrightarrow [ppERK-ERKPH]$,  
 $x_{22} \leftrightarrow [pERK-ERKPH]$,
\end{small}

we obtain these binomials that generate the steady state ideal:           

\begin{small}
\begin{tabular}{lllll}
 ${(k_2+k_3)}x_{13} -{k_1}x_1x_9$ & &
 ${(k_5+k_6)}x_{18} -{k_4}x_2x_{10}$ & & $k_3x_{13}-k_6x_{18}$\\
 ${(k_8+k_9)}x_{14} -{k_7}x_2x_3$ & & 
 ${(k_{14}+k_{15})}x_{19} -{k_{13}}x_5x_{11}$ & & $k_9x_{14}-k_{18}x_{20}$\\
 ${(k_{11}+k_{12})}x_{15} -{k_{10}}x_2x_4$ & &
 ${(k_{17}+k_{18})}x_{20} -{k_{16}}x_4x_{11}$ & & $k_{12}x_{15}-k_{15}x_{19}$\\
 ${(k_{20}+k_{21})}x_{16} -{k_{19}}x_5x_6$ & &
 ${(k_{26}+k_{27})}x_{21} -{k_{25}}x_8x_{12}$ & & $k_{21}x_{16}-k_{30}x_{22}$\\
 ${(k_{23}+k_{24})}x_{17} -{k_{22}}x_5x_7$ & & 
 ${(k_{29}+k_{30})}x_{22} -{k_{28}}x_7x_{12}$ & & $k_{24}x_{17}-k_{27}x_{21}$\\
\end{tabular}  
\end{small}
    
\medskip

We the aid of the matrices and vectors we show below, we found two steady states for 
this network. The first one, $x^1$, is approximately:

\begin{small}
\begin{tabular}{lll}
 $[RAF]=3.9412$, & & $[pRAF]=3.9412$, \\
 $[MEK]=0.0005$,& & $[pMEK]=0.0439$,\\
 $[ppMEK]=1.8241$, & &  $[ERK]=0.0084$, \\
 $[pERK]=1.1512$, & & $\textcolor{red}{[ppERK]=19.3999}$,  \\
 $[RAS]=0.4048$, & & $[RAFPH]=0.4048$,\\
 $[MEKPH]=0.0099$, & & $[ERKPH]=1.4295$ \\
 $[RAS-RAF]=1.5952$, & & $[MEK-pRAF]=0.0531$, \\
 $[pMEK-pRAF]=1.7369$, & & $[ERK-ppMEK]=0.0237$,\\
 $[pERK-ppMEK]=0.9915$,& & $[RAF-RAFPH]=1.5952$, \\
 $[ppMEK-MEKPH]=7.2949$, & & $[pMEK-MEKPH]=0.0010$,\\
 $[ppERK-ERKPH]=0.9915$, & &$[pERK-ERKPH]=0.7803$ 
\end{tabular}

\end{small}

The second steady state, $x^2$, is then built as:

\begin{small}
\begin{tabular}{lll}
 $[RAF]=0.4715$, & & $[pRAF]=0.4715$, \\
 $[MEK]=0.8619$,& & $[pMEK]=0.0126$,\\
 $[ppMEK]=0.0001$, & & $[ERK]=23.1647$, \\
 $[pERK]=0.0724$, & & $\textcolor{red}{[ppERK]=0.00003}$, \\
 $[RAS]=1.3591$, & & $[RAFPH]=1.3591$,\\
 $[MEKPH]=6.8520$, & & $[ERKPH]=3.0951$ \\
 $[RAS-RAF]=0.6409$, & & $[MEK-pRAF]=10.5784$, \\
 $[pMEK-pRAF]=0.0597$, & & $[ERK-ppMEK]=0.0032$,\\
 $[pERK-ppMEK]=0.0001$,& & $[RAF-RAFPH]=0.6409$, \\
 $[ppMEK-MEKPH]=0.2506$, & & $[pMEK-MEKPH]=0.2032$,\\
 $[ppERK-ERKPH]=0.0001$, & &$[pERK-ERKPH]=0.1062$ 
\end{tabular}
\end{small}

\medskip

Both steady states can be shown to be stable, and the total amounts defining the 
corresponding stoichiometric compatibility class are 
\begin{center}
[RAF]$_{tot}=  12.8629,$ 
[MEK]$_{tot}=11.9697,$ 
[ERK]$_{tot}=23.3465,$ 
[RAS]$_{tot}=2,$
[RAFPH]$_{tot}=2,$
[MEKPH]$_{tot}=7.3058,$ 
and [ERKPH]$_{tot}=3.2013.$
\end{center}

The rate constants that arise for the system to have the previous stable 
steady states are the following: 

\begin{small}
\begin{tabular}{lllll} 
$k_1=1.2537,$ & 
$k_2=0.6269,$ & 
$k_3=0.6269,$ & 
$k_4=1.2537,$ & 
$k_5=0.6269,$\\
$k_6=0.6269,$ & 
$k_7=5.3900,$ & 
$k_8=0.1882,$ & 
$k_9=0.0188,$ & 
$k_{10}=5.8335,$\\
$k_{11}=0.5757,$ & 
$k_{12}=0.0058,$ & 
$k_{13}=6.1030,$ & 
$k_{14}=0.0137,$ & 
$k_{15}=0.0014,$\\
$k_{16}=4.6076,$ & 
$k_{17}=0.9800,$ & 
$k_{18}=0.9800,$ & 
$k_{19}=7.1500,$ & 
$k_{20}=0.4218,$\\
$k_{21}=4.2181,$ & 
$k_{22}=0.9524,$ & 
$k_{23}=1.0086,$ & 
$k_{24}=1.0086,$ & 
$k_{25}=0.0721,$\\
$k_{26}=1.0086,$ & 
$k_{27}=1.0086,$ & 
$k_{28}=0.6684,$ & 
$k_{29}=1.2816,$ & 
$k_{30}=0.0128.$ 
\end{tabular}
\end{small}

\medskip

The matrix $A$ we chose from the binomials above can be found below.
The vectors $\alpha \in \textrm{Rowspan}(A)$ and $\sigma \in \mathcal{S}$ 
with $\textrm{sg}(\alpha_i)=\textrm{sg}(\sigma_i)$ for 
$i=1, \dots, 22$ that we found are:
\begin{small}
\begin{align*}
 \alpha=(& {\scriptstyle   -2.1233, \,   -2.1233, \,    7.4172, \,   -1.2477, \,   -9.9127, \,    
7.9181, \,   -2.7671, \,  -13.4522, \,    1.2113, \,    1.2113, \,    6.5417, \,} \\
&{\scriptstyle     0.7725, \, -0.9120, \,    5.2940, \,   -3.3710, \,   -1.9946, \,  -12.6797, \,   -0.9120, \,   -3.3710, \,   
5.2940, \,  -12.6797, \,   -1.9946}),\\
\\
\sigma=(&{\scriptstyle   -3.4697, \,  -3.4697, \,   0.8613, \,  -0.0313, \,  -1.8240, \,  23.1563, \,
  -1.0789, \, -19.3998, \,   0.9544, \,   0.9544, \,   6.8421, \,}\\
&{\scriptstyle   1.6656, \,   -0.9544, \,  10.5253, \,  -1.6772, \,  -0.0205, \,  -0.9915, \,  -0.9544, \,  -7.0443, \,   
0.2022, \,  -0.9915, \,  -0.6741}),
\end{align*}
\end{small}
where the sign pattern is 
\begin{center}
 $\text{sign}(\alpha)=\text{sign}(\sigma)=({\scriptstyle -,\,  -,\,   +,\,-,\,  -,\,  +,\,-,\, -,\,  +,\, +,\,  +,\,  
+,\, -,\,   +,\,  -,\,  -,\, -,\,  -,\,-,\,   +,\, -,\,-}).$
\end{center}

We present below the matrices $N$ and $M$, and vector $\lambda$ 
where the order for the reactions is defined by the 
subindices of the rate constants.

\begin{scriptsize}
$M=\left(\begin{array}{ccccccccccccccc}
1 & 1 & 0 & 0 & 0 & 0 & 0 & 0 & 0 & 0 & 0 & 0 & 0 & 0 & 0 \\
1 & 0 & 0 & 0 & 0 & 0 & 0 & 0 & 0 & 0 & 0 & 0 & 0 & 0 & 0 \\
0 & 1 & 0 & 0 & 0 & 0 & 0 & 0 & 0 & 0 & 0 & 0 & 0 & 0 & 0 \\
0 & 1 & 1 & 0 & 0 & 0 & 0 & 0 & 0 & 0 & 0 & 0 & 0 & 0 & 0 \\
0 & 0 & 1 & 0 & 0 & 0 & 0 & 0 & 0 & 0 & 0 & 0 & 0 & 0 & 0 \\
0 & 1 & 0 & 0 & 0 & 0 & 0 & 0 & 0 & 0 & 0 & 0 & 0 & 0 & 0 \\
0 & 0 & 0 & 1 & 1 & 0 & 0 & 0 & 0 & 0 & 0 & 0 & 0 & 0 & 0 \\
0 & 0 & 0 & 1 & 0 & 0 & 0 & 0 & 0 & 0 & 0 & 0 & 0 & 0 & 0 \\
0 & 0 & 0 & 0 & 1 & 0 & 0 & 0 & 0 & 0 & 0 & 0 & 0 & 0 & 0 \\
0 & 0 & 0 & 0 & 0 & 1 & 1 & 0 & 0 & 0 & 0 & 0 & 0 & 0 & 0 \\
0 & 0 & 0 & 0 & 0 & 1 & 0 & 0 & 0 & 0 & 0 & 0 & 0 & 0 & 0 \\
0 & 0 & 0 & 0 & 0 & 0 & 1 & 0 & 0 & 0 & 0 & 0 & 0 & 0 & 0 \\
0 & 0 & 0 & 0 & 0 & 0 & 1 & 1 & 0 & 0 & 0 & 0 & 0 & 0 & 0 \\
0 & 0 & 0 & 0 & 0 & 0 & 0 & 1 & 0 & 0 & 0 & 0 & 0 & 0 & 0 \\
0 & 0 & 0 & 0 & 0 & 0 & 1 & 0 & 0 & 0 & 0 & 0 & 0 & 0 & 0 \\
0 & 0 & 0 & 0 & 1 & 0 & 0 & 0 & 1 & 0 & 0 & 0 & 0 & 0 & 0 \\
0 & 0 & 0 & 0 & 0 & 0 & 0 & 0 & 1 & 0 & 0 & 0 & 0 & 0 & 0 \\
0 & 0 & 0 & 0 & 1 & 0 & 0 & 0 & 0 & 0 & 0 & 0 & 0 & 0 & 0 \\
0 & 0 & 0 & 0 & 0 & 0 & 0 & 0 & 0 & 1 & 1 & 0 & 0 & 0 & 0 \\
0 & 0 & 0 & 0 & 0 & 0 & 0 & 0 & 0 & 1 & 0 & 0 & 0 & 0 & 0 \\
0 & 0 & 0 & 0 & 0 & 0 & 0 & 0 & 0 & 0 & 1 & 0 & 0 & 0 & 0 \\
0 & 0 & 0 & 0 & 0 & 0 & 0 & 0 & 0 & 0 & 0 & 1 & 1 & 0 & 0 \\
0 & 0 & 0 & 0 & 0 & 0 & 0 & 0 & 0 & 0 & 0 & 1 & 0 & 0 & 0 \\
0 & 0 & 0 & 0 & 0 & 0 & 0 & 0 & 0 & 0 & 0 & 0 & 1 & 0 & 0 \\
0 & 0 & 0 & 0 & 0 & 0 & 0 & 0 & 0 & 0 & 0 & 0 & 1 & 1 & 0 \\
0 & 0 & 0 & 0 & 0 & 0 & 0 & 0 & 0 & 0 & 0 & 0 & 0 & 1 & 0 \\
0 & 0 & 0 & 0 & 0 & 0 & 0 & 0 & 0 & 0 & 0 & 0 & 1 & 0 & 0 \\
0 & 0 & 0 & 0 & 0 & 0 & 0 & 0 & 0 & 0 & 1 & 0 & 0 & 0 & 1 \\
0 & 0 & 0 & 0 & 0 & 0 & 0 & 0 & 0 & 0 & 0 & 0 & 0 & 0 & 1 \\
0 & 0 & 0 & 0 & 0 & 0 & 0 & 0 & 0 & 0 & 1 & 0 & 0 & 0 & 0 
\end{array}\right)$
\end{scriptsize}

\medskip

$\lambda=(1,1,1,0.01,0.001,1,0.01,0.1,0.001,0.01,0.1,1,1,1,1)$.

\begin{landscape}

\begin{footnotesize}
\noindent $A=\left(\begin{array}{rrrrrrrrrrrrrrrrrrrrrr}
-1 & 0 & 0 & 0 & 0 & 0 & 0 & 0 & 1 & 0 & 0 & 0 & 0 & 0 & 0 & 0 & 0 & 0 & 0 & 0 & 0 & 0\\
 0 & 1 &-1 &-1 &-1 & 2 & 1 & 0 & 0 &-1 & 1 & 0 & 0 & 0 & 0 & 1 & 0 & 0 & 0 & 0 & 0 & 1\\
 0 & 0 & 1 & 0 &-1 & 2 & 1 & 0 & 0 & 0 & 1 & 0 & 0 & 1 & 0 & 1 & 0 & 0 & 0 & 1 & 0 & 1\\
 0 & 0 & 0 & 1 & 2 &-4 &-2 & 0 & 0 & 0 &-1 & 0 & 0 & 0 & 1 &-2 & 0 & 0 & 1 & 0 & 0 &-2\\
 0 & 0 & 0 & 0 & 0 & 1 & 0 &-1 & 0 & 0 & 0 & 1 & 0 & 0 & 0 & 1 & 0 & 0 & 0 & 0 & 0 & 1\\
 0 & 0 & 0 & 0 & 0 & 0 & 1 & 2 & 0 & 0 & 0 &-1 & 0 & 0 & 0 & 0 & 1 & 0 & 0 & 0 & 1 & 0\\
 0 & 0 & 0 & 0 & 0 & 0 & 0 & 0 & 1 & 1 & 0 & 0 & 1 & 0 & 0 & 0 & 0 & 1 & 0 & 0 & 0 & 0
   \end{array}\right)$
\end{footnotesize}

\bigskip

\resizebox{\linewidth}{!}{%
$N=\left(
\begin{array}{cccccccccccccccccccccccccccccc}
-1& 1& 0& 0& 0& 1&0&0&0&0&0&0&0&0&0&0&0&0&0&0&0&0&0&0&0&0&0&0&0&0\\
 0& 0& 1&-1& 1& 0&-1&1&1&-1&1&1&0&0&0&0&0&0&0&0&0&0&0&0&0&0&0&0&0&0\\
 0& 0& 0& 0& 0& 0&-1&1&0&0&0&0&0&0&0&0&0&1&0&0&0&0&0&0&0&0&0&0&0&0\\
 0& 0& 0& 0& 0& 0&0&0&1&-1&1&0&0&0&1&-1&1&0&0&0&0&0&0&0&0&0&0&0&0&0\\
 0& 0& 0& 0& 0& 0&0&0&0&0&0&1&-1&1&0&0&0&0&-1&1&1&-1&1&1&0&0&0&0&0&0\\
 0& 0& 0& 0& 0& 0&0&0&0&0&0&0&0&0&0&0&0&0&-1&1&0&0&0&0&0&0&0&0&0&1\\
 0& 0& 0& 0& 0& 0&0&0&0&0&0&0&0&0&0&0&0&0&0&0&1&-1&1&0&0&0&1&-1&1&0\\
 0& 0& 0& 0& 0& 0&0&0&0&0&0&0&0&0&0&0&0&0&0&0&0&0&0&1&-1&1&0&0&0&0\\
-1& 1& 1& 0& 0& 0&0&0&0&0&0&0&0&0&0&0&0&0&0&0&0&0&0&0&0&0&0&0&0&0\\
 0& 0& 0&-1& 1& 1&0&0&0&0&0&0&0&0&0&0&0&0&0&0&0&0&0&0&0&0&0&0&0&0\\
 0& 0& 0& 0& 0& 0&0&0&0&0&0&0&-1&1&1&-1&1&1&0&0&0&0&0&0&0&0&0&0&0&0\\
 0& 0& 0& 0& 0& 0&0&0&0&0&0&0&0&0&0&0&0&0&0&0&0&0&0&0&-1&1&1&-1&1&1\\
 1&-1&-1& 0& 0& 0&0&0&0&0&0&0&0&0&0&0&0&0&0&0&0&0&0&0&0&0&0&0&0&0\\
 0& 0& 0& 0& 0& 0&1&-1&-1&0&0&0&0&0&0&0&0&0&0&0&0&0&0&0&0&0&0&0&0&0\\
 0& 0& 0& 0& 0& 0&0&0&0&1&-1&-1&0&0&0&0&0&0&0&0&0&0&0&0&0&0&0&0&0&0\\
 0& 0& 0& 0& 0& 0&0&0&0&0&0&0&0&0&0&0&0&0&1&-1&-1&0&0&0&0&0&0&0&0&0\\
 0& 0& 0& 0& 0& 0&0&0&0&0&0&0&0&0&0&0&0&0&0&0&0&1&-1&-1&0&0&0&0&0&0\\
 0& 0& 0& 1&-1&-1&0&0&0&0&0&0&0&0&0&0&0&0&0&0&0&0&0&0&0&0&0&0&0&0\\
 0& 0& 0& 0& 0& 0&0&0&0&0&0&0&1&-1&-1&0&0&0&0&0&0&0&0&0&0&0&0&0&0&0\\
 0& 0& 0& 0& 0& 0&0&0&0&0&0&0&0&0&0&1&-1&-1&0&0&0&0&0&0&0&0&0&0&0&0\\
 0& 0& 0& 0& 0& 0&0&0&0&0&0&0&0&0&0&0&0&0&0&0&0&0&0&0&1&-1&-1&0&0&0\\
 0& 0& 0& 0& 0& 0&0&0&0&0&0&0&0&0&0&0&0&0&0&0&0&0&0&0&0&0&0&1&-1&-1
 \end{array}\right)$
 }

\end{landscape}

\section{The ERK network without feedback and two phosphatases.}

We present in this appendix the matrices, vectors, constants and corresponding 
(stable) steady states that prove the capacity for multistationarity 
for the system without feedback and two phosphatases, presented in 
Subsection~\ref{subsec:red2}. 

\medskip

The conservation relations of this system are:

\medskip

\begin{small}
\begin{align*}
 \text{[RAF]+[pRAF]+[RAS-RAF]+[MEK-pRAF]+[pMEK-pRAF]+}&\\
 \text{+[pRAF-RAFPH]}&=C_1\\ 
\text{[MEK]+[pMEK]+[ppMEK]+[MEK-pRAF]+[pMEK-pRAF]+}&\\
\text{+[ERK-ppMEK]+[pERK-ppMEK]+[ppMEK-PH]+[pMEK-PH]}&=C_2\\ 
\text{[ERK]+[pERK]+[ppERK]+[ERK-ppMEK]+[pERK-ppMEK]+}&\\
\text{+[ppERK-PH]+[pERK-PH]}&=C_3\\ 
\text{[RAS]+[RAS-RAF]}&=C_4\\ 
\text{[RAFPH]+[RAF-RAFPH]}&=C_5\\ 
\text{[PH]+[ppMEK-PH]+[pMEK-PH]+[ppERK-PH]+[pERK-PH]}&=C_6
\end{align*}
\end{small}

\medskip

Under mass-action kinetics, the steady state ideal for this network is binomial. In fact, if we consider
the following order of the species:

\begin{small}
 $x_1 \leftrightarrow [RAF]$,  $x_2 \leftrightarrow [pRAF]$,
 $x_3 \leftrightarrow [MEK]$, $x_4 \leftrightarrow [pMEK]$, $x_5 \leftrightarrow [ppMEK]$,
 
 $x_6 \leftrightarrow [ERK]$,  $x_7 \leftrightarrow [pERK]$,  $x_8 \leftrightarrow [ppERK]$,
 $x_9 \leftrightarrow [RAS]$, $x_{10} \leftrightarrow [RAFPH]$, 
 
 $x_{11} \leftrightarrow [PH]$,   $x_{12} \leftrightarrow [RAS-RAF]$,  $x_{13} \leftrightarrow [MEK-pRAF]$,  
 
 $x_{14} \leftrightarrow [pMEK-pRAF]$, 
 $x_{15} \leftrightarrow [ERK-ppMEK]$,  $x_{16} \leftrightarrow [pERK-ppMEK]$,

 $x_{17} \leftrightarrow [RAF-RAFPH]$,  $x_{18} \leftrightarrow [ppMEK-PH]$,  
 $x_{19} \leftrightarrow [pMEK-PH]$,  
 
 $x_{20} \leftrightarrow [ppERK-PH]$,  $x_{21} \leftrightarrow [pERK-PH]$,
\end{small}

we obtain these binomials that generate the steady state ideal:           

\begin{small}
\begin{tabular}{lllll}
 ${(k_{2}+k_3)}x_{12} -{k_1}x_1x_9$ & &
 ${(k_{5}+k_6)}x_{17} -{k_4}x_2x_{10}$ & & $k_3x_{12}-k_6x_{17}$\\
 ${(k_{8}+k_9)}x_{13} -{k_7}x_2x_3$ & & 
 ${(k_{14}+k_{15})}x_{18} -{k_{13}}x_5x_{11}$ & & $k_9x_{13}-k_{18}x_{19}$\\
 ${(k_{11}+k_{12})}x_{14} -{k_{10}}x_2x_4$ & &
 ${(k_{17}+k_{18})}x_{19} -{k_{16}}x_4x_{11}$ & & $k_{12}x_{14}-k_{15}x_{18}$\\
 ${(k_{20}+k_{21})}x_{15} -{k_{19}}x_5x_6$ & &
 ${(k_{26}+k_{27})}x_{20} -{k_{25}}x_8x_{11}$ & & $k_{21}x_{15}-k_{30}x_{21}$\\
 ${(k_{23}+k_{24})}x_{16} -{k_{22}}x_5x_7$ & & 
 ${(k_{29}+k_{30})}x_{21} -{k_{28}}x_7x_{11}$ & & $k_{24}x_{16}-k_{27}x_{20}$\\
\end{tabular}  
\end{small}
    
\bigskip

We the aid of the matrices and vectors we show below, we found two steady states for 
this network. The first one, $x^1$, is approximately:

\begin{small}
\begin{tabular}{lll}
 $[RAF]=1$, & & $[pRAF]=1$, \\
 $[MEK]=0.0939$,& & $[pMEK]=0.1582$,\\
 $[ppMEK]=0.1019$, & &  $[ERK]=0$, \\
 $[pERK]=0.0116$, & & $\textcolor{red}{[ppERK]=3.5988}$,  \\
 $[RAS]=1$, & & $[RAFPH]=1$, \\
 $[PH]0.0210=$, &  & $[RAS-RAF]=1$,\\
 $[MEK-pRAF]=0.0157$, & & $[pMEK-pRAF]=0.1582$, \\
 $[ERK-ppMEK]=0$, & & $[pERK-ppMEK]=0.0816$,\\
 $[RAF-RAFPH]=1$, & & $[ppMEK-PH]=1.2656$,\\
 $[pMEK-PH]=0.0313$, & & $[ppERK-PH]=0.0816$, \\
 $[pERK-PH]=0$&&
\end{tabular}
\end{small}

\bigskip

The second steady state, $x^2$, would then be:

\begin{small}
\begin{tabular}{lll}
 $[RAF]=1$, & & $[pRAF]=1$, \\
 $[MEK]=0.6939$,& & $[pMEK]=0.0582$,\\
 $[ppMEK]=0.0019$, & & $[ERK]=1.4000$, \\
 $[pERK]=1.4116$,& & $\textcolor{red}{[ppERK]=0.3988}$, \\
 $[RAS]=1$, & & $[RAFPH]=1$, \\
 $[PH]=0.4210$, &  & $[RAS-RAF]=1$, \\
 $[MEK-pRAF]=0.1157$, & & $[pMEK-pRAF]=0.0582$, \\
 $[ERK-ppMEK]=0.1$, & & $[pERK-ppMEK]=0.1816$,\\
 $[RAF-RAFPH]=1$, & & $[ppMEK-MEKPH]=0.4656$,\\
 $[pMEK-MEKPH]=0.2313$, &&  $[ppERK-ERKPH]=0.1816$,\\
 $[pERK-ERKPH]=0.1$&&  
\end{tabular}
\end{small}

Both steady states can be shown to be stable, and the total amounts defining the 
corresponding stoichiometric compatibility class are 
\begin{center}
[RAF]$_{tot}= 4.1738,$ 
[MEK]$_{tot}=1.9063,$ 
[ERK]$_{tot}=3.7737,$ 
[RAS]$_{tot}=2,$
[RAFPH]$_{tot}=2,$
and [PH]$_{tot}=1.3995.$
\end{center}

The rate constants that arise for the system to have the previous stable 
steady states are the following:

\begin{small}
\begin{tabular}{llll}
$k_1=2,$ & 
$k_{2}=1,$ & 
$k_3=1,$ & 
$k_{4}=2,$ \\
$k_{5}=1,$ &
$k_{6}=1,$ & 
$k_{7}=1.1713,$ & 
$k_{8}=6.3891,$ \\
$k_{9}=0.6389,$ & 
$k_{10}=6.9533,$ &
$k_{11}=6.3212,$ & 
$k_{12}=0.6321,$ \\
$k_{13}=93.6799,$ & 
$k_{14}=0.0790,$ & 
$k_{15}=0.0790,$ &
$k_{16}=6.0322,$ \\
$k_{17}=0.3195,$ & 
$k_{18}=0.3195,$ & 
$k_{19}=18.6873,$ & 
$k_{20}=0.2440,$\\
$k_{21}=0.2440,$ & 
$k_{22}=16.9005,$ & 
$k_{23}=0.1226,$ & 
$k_{24}=0.1226,$ \\
$k_{25}=13.3910,$ &
$k_{26}=12.2554,$ & 
$k_{27}=0.1226,$ & 
$k_{28}=4.1482,$ \\
$k_{29}=24.3960,$ & 
$k_{30}=0.2440.$ & &
\end{tabular}
\end{small}

\medskip

The matrix $A$ we chose from the binomials above is depicted below.
The vectors $\alpha \in \textrm{Rowspan}(A)$ and $\sigma \in \mathcal{S}$ 
with $\textrm{sg}(\alpha_i)=\textrm{sg}(\sigma_i)$ for $i=1, \dots, 21$ that we found are

\begin{align*}
\alpha=&({\scriptstyle 0,\; 0,\; 2,\; -1,\; -4,\; 11.8,\; 4.8,\; -2.2,\; 0,\;
0,\; 3,\; 0,\; 2,\; -1,\; 7.8,\; 0.8,\; 0,\; -1,\; 2,\; 0.8,\; 7.8}),\\
\sigma=&({\scriptstyle 0,\; 0,\; 0.6,\; -0.1,\; -0.1,\; 1.4, 1.4,\; -3.2,\; 0,\; 0,\;
0.4,\; 0,\; 0.1,\; -0.1,\; 0.1,\; 0.1,\; 0,\; -0.8,\; 0.2,\; 0.1,\; 0.1}),
\end{align*}
where the sign pattern is 

\begin{center}
 $\text{sign}(\alpha)=\text{sign}(\sigma)=({\scriptstyle 0,\,  0,\,   +,\,-,\,  -,\,  +,\,+,\, -,\,  0,\, 0,\,  +,\,  
0,\, +,\,   -,\,  +,\,  +,\, 0,\,  -,\,+,\,   +,\, +})$.
\end{center}
 
\medskip

We present below the matrices $N$ and $M$, and vector $\lambda$ 
where the order for the reactions is defined by the 
subindices of the rate constants.

\medskip

\begin{scriptsize}
$M=\left(\begin{array}{ccccccccccccccc}
1 & 0 & 0 & 0 & 0 & 0 & 0 & 0 & 0 & 0 & 1 & 0 & 0 & 0 & 0\\
1 & 0 & 0 & 0 & 0 & 0 & 0 & 0 & 0 & 0 & 0 & 0 & 0 & 0 & 0\\
0 & 0 & 0 & 0 & 0 & 0 & 0 & 0 & 0 & 0 & 1 & 0 & 0 & 0 & 0\\
0 & 1 & 0 & 0 & 0 & 0 & 0 & 0 & 0 & 0 & 1 & 0 & 0 & 0 & 0\\
0 & 1 & 0 & 0 & 0 & 0 & 0 & 0 & 0 & 0 & 0 & 0 & 0 & 0 & 0\\
0 & 0 & 0 & 0 & 0 & 0 & 0 & 0 & 0 & 0 & 1 & 0 & 0 & 0 & 0\\
0 & 0 & 0 & 0 & 1 & 0 & 0 & 0 & 0 & 0 & 0 & 1 & 0 & 0 & 0\\
0 & 0 & 0 & 0 & 1 & 0 & 0 & 0 & 0 & 0 & 0 & 0 & 0 & 0 & 0\\
0 & 0 & 0 & 0 & 0 & 0 & 0 & 0 & 0 & 0 & 0 & 1 & 0 & 0 & 0\\
0 & 0 & 1 & 0 & 0 & 0 & 0 & 0 & 0 & 0 & 0 & 0 & 1 & 0 & 0\\
0 & 0 & 1 & 0 & 0 & 0 & 0 & 0 & 0 & 0 & 0 & 0 & 0 & 0 & 0\\
0 & 0 & 0 & 0 & 0 & 0 & 0 & 0 & 0 & 0 & 0 & 0 & 1 & 0 & 0\\
0 & 0 & 0 & 1 & 0 & 0 & 0 & 0 & 0 & 0 & 0 & 0 & 1 & 0 & 0\\
0 & 0 & 0 & 1 & 0 & 0 & 0 & 0 & 0 & 0 & 0 & 0 & 0 & 0 & 0\\
0 & 0 & 0 & 0 & 0 & 0 & 0 & 0 & 0 & 0 & 0 & 0 & 1 & 0 & 0\\
0 & 0 & 0 & 0 & 0 & 1 & 0 & 0 & 0 & 0 & 0 & 1 & 0 & 0 & 0\\
0 & 0 & 0 & 0 & 0 & 1 & 0 & 0 & 0 & 0 & 0 & 0 & 0 & 0 & 0\\
0 & 0 & 0 & 0 & 0 & 0 & 0 & 0 & 0 & 0 & 0 & 1 & 0 & 0 & 0\\
0 & 0 & 0 & 0 & 0 & 0 & 1 & 0 & 0 & 0 & 0 & 0 & 0 & 0 & 1\\
0 & 0 & 0 & 0 & 0 & 0 & 1 & 0 & 0 & 0 & 0 & 0 & 0 & 0 & 0\\
0 & 0 & 0 & 0 & 0 & 0 & 0 & 0 & 0 & 0 & 0 & 0 & 0 & 0 & 1\\
0 & 0 & 0 & 0 & 0 & 0 & 0 & 1 & 0 & 0 & 0 & 0 & 0 & 1 & 0\\
0 & 0 & 0 & 0 & 0 & 0 & 0 & 1 & 0 & 0 & 0 & 0 & 0 & 0 & 0\\
0 & 0 & 0 & 0 & 0 & 0 & 0 & 0 & 0 & 0 & 0 & 0 & 0 & 1 & 0\\
0 & 0 & 0 & 0 & 0 & 0 & 0 & 0 & 1 & 0 & 0 & 0 & 0 & 1 & 0\\
0 & 0 & 0 & 0 & 0 & 0 & 0 & 0 & 1 & 0 & 0 & 0 & 0 & 0 & 0\\
0 & 0 & 0 & 0 & 0 & 0 & 0 & 0 & 0 & 0 & 0 & 0 & 0 & 1 & 0\\
0 & 0 & 0 & 0 & 0 & 0 & 0 & 0 & 0 & 1 & 0 & 0 & 0 & 0 & 1\\
0 & 0 & 0 & 0 & 0 & 0 & 0 & 0 & 0 & 1 & 0 & 0 & 0 & 0 & 0\\
0 & 0 & 0 & 0 & 0 & 0 & 0 & 0 & 0 & 0 & 0 & 0 & 0 & 0 & 1
\end{array}\right)$
\end{scriptsize}

\medskip

$\lambda=(1,1,1,0.1,0.1,0.01,0.00001,0.01,1,0.001,1,0.01,0.1,0.01,0.00001)$.

\begin{landscape}
 \noindent {\footnotesize $A=\left(\begin{array}{rrrrrrrrrrrrrrrrrrrrr}
-1 & 0 & 0 & 0 & 0 & 0 & 0 & 0 & 1 & 0 & 0 & 0 & 0 & 0 & 0 & 0 & 0 & 0 & 0 & 0 & 0 \\
 0 & 1 &-1 &-1 &-1 & 3 & 1 &-1 & 0 &-1 & 1 & 0 & 0 & 0 & 2 & 0 & 0 & 0 & 0 & 0 & 2 \\
 0 & 0 & 0 & 0 & 0 & 1 & 1 & 1 & 0 & 0 & 0 & 0 & 0 & 0 & 1 & 1 & 0 & 0 & 0 & 1 & 1 \\
 1 & 0 & 0 & 0 & 0 & 0 & 0 & 0 & 0 & 1 & 0 & 1 & 0 & 0 & 0 & 0 & 1 & 0 & 0 & 0 & 0 \\
 0 & 1 &-1 & 0 & 1 &-2 &-1 & 0 & 0 &-1 & 0 & 0 & 0 & 1 &-1 & 0 & 0 & 1 & 0 & 0 &-1 \\
 0 &-1 & 2 & 1 & 0 & 0 & 0 & 0 & 0 & 1 & 0 & 0 & 1 & 0 & 0 & 0 & 0 & 0 & 1 & 0 & 0 \\
   \end{array}\right)$}
   
   \bigskip
   
\resizebox{\linewidth}{!}{%
$N=\left(
\begin{array}{cccccccccccccccccccccccccccccc}
 -1 & 1 & 0 & 0 & 0 & 1 & 0 & 0 & 0 & 0 & 0 & 0 & 0 & 0 & 0 & 0 & 0 & 0 & 0 & 0 & 0 & 0 & 0 & 0 & 0 & 0 & 0 & 0 & 0 & 0 \\
  0 & 0 & 1 &-1 & 1 & 0 &-1 & 1 & 1 &-1 & 1 & 1 & 0 & 0 & 0 & 0 & 0 & 0 & 0 & 0 & 0 & 0 & 0 & 0 & 0 & 0 & 0 & 0 & 0 & 0 \\
  0 & 0 & 0 & 0 & 0 & 0 &-1 & 1 & 0 & 0 & 0 & 0 & 0 & 0 & 0 & 0 & 0 & 1 & 0 & 0 & 0 & 0 & 0 & 0 & 0 & 0 & 0 & 0 & 0 & 0 \\
  0 & 0 & 0 & 0 & 0 & 0 & 0 & 0 & 1 &-1 & 1 & 0 & 0 & 0 & 1 &-1 & 1 & 0 & 0 & 0 & 0 & 0 & 0 & 0 & 0 & 0 & 0 & 0 & 0 & 0 \\
  0 & 0 & 0 & 0 & 0 & 0 & 0 & 0 & 0 & 0 & 0 & 1 &-1 & 1 & 0 & 0 & 0 & 0 &-1 & 1 & 1 &-1 & 1 & 1 & 0 & 0 & 0 & 0 & 0 & 0 \\
  0 & 0 & 0 & 0 & 0 & 0 & 0 & 0 & 0 & 0 & 0 & 0 & 0 & 0 & 0 & 0 & 0 & 0 &-1 & 1 & 0 & 0 & 0 & 0 & 0 & 0 & 0 & 0 & 0 & 1 \\
  0 & 0 & 0 & 0 & 0 & 0 & 0 & 0 & 0 & 0 & 0 & 0 & 0 & 0 & 0 & 0 & 0 & 0 & 0 & 0 & 1 &-1 & 1 & 0 & 0 & 0 & 1 &-1 & 1 & 0 \\
  0 & 0 & 0 & 0 & 0 & 0 & 0 & 0 & 0 & 0 & 0 & 0 & 0 & 0 & 0 & 0 & 0 & 0 & 0 & 0 & 0 & 0 & 0 & 1 &-1 & 1 & 0 & 0 & 0 & 0 \\
 -1 & 1 & 1 & 0 & 0 & 0 & 0 & 0 & 0 & 0 & 0 & 0 & 0 & 0 & 0 & 0 & 0 & 0 & 0 & 0 & 0 & 0 & 0 & 0 & 0 & 0 & 0 & 0 & 0 & 0 \\
  0 & 0 & 0 &-1 & 1 & 1 & 0 & 0 & 0 & 0 & 0 & 0 & 0 & 0 & 0 & 0 & 0 & 0 & 0 & 0 & 0 & 0 & 0 & 0 & 0 & 0 & 0 & 0 & 0 & 0 \\
  0 & 0 & 0 & 0 & 0 & 0 & 0 & 0 & 0 & 0 & 0 & 0 &-1 & 1 & 1 &-1 & 1 & 1 & 0 & 0 & 0 & 0 & 0 & 0 &-1 & 1 & 1 &-1 & 1 & 1 \\
  1 &-1 &-1 & 0 & 0 & 0 & 0 & 0 & 0 & 0 & 0 & 0 & 0 & 0 & 0 & 0 & 0 & 0 & 0 & 0 & 0 & 0 & 0 & 0 & 0 & 0 & 0 & 0 & 0 & 0 \\
  0 & 0 & 0 & 0 & 0 & 0 & 1 &-1 &-1 & 0 & 0 & 0 & 0 & 0 & 0 & 0 & 0 & 0 & 0 & 0 & 0 & 0 & 0 & 0 & 0 & 0 & 0 & 0 & 0 & 0 \\
  0 & 0 & 0 & 0 & 0 & 0 & 0 & 0 & 0 & 1 &-1 &-1 & 0 & 0 & 0 & 0 & 0 & 0 & 0 & 0 & 0 & 0 & 0 & 0 & 0 & 0 & 0 & 0 & 0 & 0 \\
  0 & 0 & 0 & 0 & 0 & 0 & 0 & 0 & 0 & 0 & 0 & 0 & 0 & 0 & 0 & 0 & 0 & 0 & 1 &-1 &-1 & 0 & 0 & 0 & 0 & 0 & 0 & 0 & 0 & 0 \\
  0 & 0 & 0 & 0 & 0 & 0 & 0 & 0 & 0 & 0 & 0 & 0 & 0 & 0 & 0 & 0 & 0 & 0 & 0 & 0 & 0 & 1 &-1 &-1 & 0 & 0 & 0 & 0 & 0 & 0 \\
  0 & 0 & 0 & 1 &-1 &-1 & 0 & 0 & 0 & 0 & 0 & 0 & 0 & 0 & 0 & 0 & 0 & 0 & 0 & 0 & 0 & 0 & 0 & 0 & 0 & 0 & 0 & 0 & 0 & 0 \\
  0 & 0 & 0 & 0 & 0 & 0 & 0 & 0 & 0 & 0 & 0 & 0 & 1 &-1 &-1 & 0 & 0 & 0 & 0 & 0 & 0 & 0 & 0 & 0 & 0 & 0 & 0 & 0 & 0 & 0 \\
  0 & 0 & 0 & 0 & 0 & 0 & 0 & 0 & 0 & 0 & 0 & 0 & 0 & 0 & 0 & 1 &-1 &-1 & 0 & 0 & 0 & 0 & 0 & 0 & 0 & 0 & 0 & 0 & 0 & 0 \\
  0 & 0 & 0 & 0 & 0 & 0 & 0 & 0 & 0 & 0 & 0 & 0 & 0 & 0 & 0 & 0 & 0 & 0 & 0 & 0 & 0 & 0 & 0 & 0 & 1 &-1 &-1 & 0 & 0 & 0 \\
  0 & 0 & 0 & 0 & 0 & 0 & 0 & 0 & 0 & 0 & 0 & 0 & 0 & 0 & 0 & 0 & 0 & 0 & 0 & 0 & 0 & 0 & 0 & 0 & 0 & 0 & 0 & 1 &-1 &-1 \\
 \end{array}\right)$
 }
   
\end{landscape}

\section{The ERK network with feedback.}
We present in this appendix the matrices, vectors, constants and corresponding 
(stable) steady states that prove the capacity for multistationarity 
for the system with a negative feedback loop, presented in 
Subsection~\ref{subsec:red3}. 

\medskip

The conservation relations of this system are:

\medskip

\begin{small}
\begin{align*}
 \text{[RAF]+[pRAF]+[RAS-RAF]+[MEK-pRAF]+[pMEK-pRAF]+}&\\
 \text{+[pRAF-RAFPH]+[pRAF-ppERK]+[Z-PH2]+[Z]}&=C_1\\ 
\text{[MEK]+[pMEK]+[ppMEK]+[MEK-pRAF]+[pMEK-pRAF]+}&\\
\text{+[ERK-ppMEK]+[pERK-ppMEK]+[ppMEK-PH]+[pMEK-PH]}&=C_2\\ 
\text{[ERK]+[pERK]+[ppERK]+[ERK-ppMEK]+[pERK-ppMEK]+}&\\
\text{+[ppERK-PH]+[pERK-PH]+[pRAF-ppERK]}&=C_3\\ 
\text{[RAS]+[RAS-RAF]}&=C_4\\ 
\text{[RAFPH]+[RAF-RAFPH]}&=C_5\\ 
\text{[PH]+[ppMEK-PH]+[pMEK-PH]+[ppERK-PH]+[pERK-PH]}&=C_6\\ 
\text{[PH2]+[Z-PH2]}&=C_7
\end{align*}
\end{small}

\bigskip

Under mass-action kinetics, the steady state ideal for this network is binomial. In fact, if we consider
the following order of the species:

\bigskip

\begin{small}
 $x_1 \leftrightarrow [RAF]$,  $x_2 \leftrightarrow [pRAF]$,
 $x_3 \leftrightarrow [MEK]$, $x_4 \leftrightarrow [pMEK]$, $x_5 \leftrightarrow [ppMEK]$,
 
 $x_6 \leftrightarrow [ERK]$,  $x_7 \leftrightarrow [pERK]$,  $x_8 \leftrightarrow [ppERK]$,
 $x_9 \leftrightarrow [RAS]$, $x_{10} \leftrightarrow [RAFPH]$, 
 
 $x_{11} \leftrightarrow [PH]$,  $x_{12} \leftrightarrow [PH2]$  
 $x_{13} \leftrightarrow [RAS-RAF]$,  $x_{14} \leftrightarrow [MEK-pRAF]$,  
 
 $x_{15} \leftrightarrow [pMEK-pRAF]$, 
 $x_{16} \leftrightarrow [ERK-ppMEK]$,  $x_{17} \leftrightarrow [pERK-ppMEK]$,
 
 $x_{18} \leftrightarrow [RAF-RAFPH]$,  $x_{19} \leftrightarrow [ppMEK-MEKPH]$,  
 
 $x_{20} \leftrightarrow [pMEK-MEKPH]$,  $x_{21} \leftrightarrow [ppERK-ERKPH]$,  
 
 $x_{22} \leftrightarrow [pERK-ERKPH]$, $x_{23} \leftrightarrow [pRAF-ppERK]$,  
 
 $x_{24} \leftrightarrow [Z-PH2]$, $x_{25} \leftrightarrow [Z]$,
\end{small}
 
 \bigskip
 
we obtain these binomials that generate the steady state ideal:           

\bigskip

\begin{small}
\begin{tabular}{lllll}
 ${(k_{2}+k_{3})}x_{13} -{k_1}x_1x_9$ & &
 ${(k_{5}+k_{6})}x_{18} -{k_{4}}x_2x_{10}$ & & $k_{3}x_{13}-k_{6}x_{18}$\\
 ${(k_{8}+k_{9})}x_{14} -{k_{7}}x_2x_3$ & & 
 ${(k_{14}+k_{15})}x_{19} -{k_{13}}x_5x_{11}$ & & $k_{9}x_{14}-k_{18}x_{20}$\\
 ${(k_{11}+k_{12})}x_{15} -{k_{10}}x_2x_4$ & &
 ${(k_{17}+k_{18})}x_{20} -{k_{16}}x_4x_{11}$ & & $k_{12}x_{15}-k_{15}x_{19}$\\
 ${(k_{20}+k_{21})}x_{16} -{k_{19}}x_5x_6$ & &
 ${(k_{26}+k_{27})}x_{21} -{k_{25}}x_8x_{11}$ & & $k_{21}x_{16}-k_{30}x_{22}$\\
 ${(k_{23}+k_{24})}x_{17} -{k_{22}}x_5x_7$ & & 
 ${(k_{29}+k_{30})}x_{22} -{k_{28}}x_7x_{11}$ & & $k_{24}x_{17}-k_{27}x_{21}$\\
 $(k_{32}+k_{33})x_{23}-k_{31}x_2x_8$ & & 
 $(k_{35}+k_{36})x_{24}-k_{34}x_{25}x_{12}$ & &  $k_{33}x_{23}-k_{36}x_{24}$\\
\end{tabular} 
\end{small}
    
\bigskip

With the aid of the matrices and vectors we show below, we found two steady states for 
this network. The first one, $x^1$, is approximately:

\bigskip

\begin{small}
\begin{tabular}{lll}
 $[RAF]=0.4723$, & & $[pRAF]=0.4723$,  \\
 $[MEK]=0.8601$,& & $[pMEK]=0.0128$, \\
 $[ppMEK]=0.0001$, & &  $[ERK]=23.1985$, \\
 $[pERK]=0.0345 $, & & $\textcolor{red}{[ppERK]=0.00001}$, \\
 $[RAS]=1.3584$, & & $[RAFPH]=1.3584$,\\ 
 $[PH]=6.7373$ & & $[PH2]=3.2013$\\
 $[RAS-RAF]=0.6416$, & & $[MEK-pRAF]=10.5743$,\\
 $[pMEK-pRAF]=0.0608$, & &  $[ERK-ppMEK]=0.0033$, \\
 $[pERK-ppMEK]=0.000002$,&& $[RAF-RAFPH]=0.6416$,\\ 
 $[ppMEK-PH]=0.2552$, & & $[pMEK-PH]=0.2031$, \\
  $[ppERK-PH]=0.000002$, & & $[pERK-PH]=0.1101$, \\
  $[pRAF-ppERK]=0.00000003$, & &$[Z-PH2]=0.0000003$,\\
  $[Z]=0.0000001$ & &
\end{tabular}
\end{small}

\bigskip

The second steady state, $x^2$, would then be:

\bigskip

\begin{small}
\begin{tabular}{lll}
 $[RAF]= 1.0761$, & & $[pRAF]=1.0761$,  \\
 $[MEK]=0.0048$, & & $[pMEK]=0.1608$,\\
 $[ppMEK]=2.6347$, & &  $[ERK]=0.0000001$,\\
 $[pERK]=0.0045$, & & $\textcolor{red}{[ppERK]=23.0822}$, \\
 $[RAS]=0.9633$, & & $[RAFPH]=0.9633$, \\
 $[PH]=0.0068$ & & $[PH2]=0.7111$\\
 $[RAS-RAF]=1.0367$, & & $[MEK-pRAF]=0.1346$,\\
 $[pMEK-pRAF]=1.7359$, & &  $[ERK-ppMEK]=0.0000004$,\\
 $[pERK-ppMEK]=0.0056$, && $[RAF-RAFPH]=1.0367$, \\ 
 $[ppMEK-PH]=7.2907$, & & $[pMEK-PH]=0.0026$, \\
 $[ppERK-PH]=0.0056$, & & $[pERK-PH]=0.00001$,\\
 $[pRAF-ppERK]=0.2485$,& &$[Z-PH2]=2.4902$,\\
 $[Z]=4.0283$ &&
\end{tabular}
\end{small}

\vspace{1.5cm}

Both steady states can be shown to be stable, and the total amounts defining the 
corresponding stoichiometric compatibility class are 

\bigskip

\begin{center}
[RAF]$_{tot}= 12.8629,$ 
[MEK]$_{tot}=11.9697,$ 
[ERK]$_{tot}=23.3465,$ 
[RAS]$_{tot}=2,$
[RAFPH]$_{tot}=2,$
[PH]$_{tot}=7.3058,$
and [PH2]$_{tot}=3.2013.$
\end{center}

\vspace{1.5cm}

The rate constants that arise for the system to have the previous stable 
steady states are the following:

\medskip

\begin{small}
$
\begin{array}{lllll}
k_1=0.0312, &
k_{2}=0.0156, &
k_3=0.0156,&
k_{4}=0.0312,&
k_{5}=0.0156,\\
k_{6}=0.0156,&
k_{7}=0.0492,&
k_{8}=0.0009,&
k_{9}=0.0009,&
k_{10}=1.8160,\\
k_{11}=0.1646,&
k_{12}=0.0165,&
k_{13}=3.1717,&
k_{14}=0.0039,&
k_{15}=0.0039,\\
k_{16}=0.2315,&
k_{17}=0.0492,&
k_{18}=0.0492,&
k_{19}=5.0662,&
k_{20}=2.9888,\\
k_{21}=0.2989,&
k_{22}=6.1986,&
k_{23}=6.5645,&
k_{24}=6.5645,&
k_{25}=0.4694,\\
k_{26}=6.5645,&
k_{27}=6.5645,&
k_{28}=0.0474,&
k_{29}=0.0908,&
k_{30}=0.0091,\\
k_{31}=0.0670,&
k_{32}=3.3460,&
k_{33}=3.3460,&
k_{34}=3.1930,&
k_{35}=3.3392,\\
k_{36}=0.3339.&&&&
\end{array}
$
\end{small}

\bigskip

Below we can find the matrix $A$ we chose from the binomials above.
The vectors $\alpha \in \textrm{Rowspan}(A)$ and $\sigma \in \mathcal{S}$ 
with $\textrm{sg}(\alpha_i)=\textrm{sg}(\sigma_i)$ for $i=1, \dots, 25$ that we found are

 \begin{small}
\begin{align*}
\alpha=(&{\scriptstyle 0.8234, \,   0.8234, \,  -5.1877, \,
2.5288, \,  10.2453, \, -19.1666, \,
-2.0282, \,  15.1102, \,  -0.3436, \,
-0.3436, \,  -6.8931, \,}\\
&{\scriptstyle  -1.5045, \, 0.4798, \,  -4.3642, \,   3.3523, \,
-8.9213, \,   8.2171, \,   0.4798, \,
3.3523, \,  -4.3642, \,   8.2171, \,
-8.9213, \,  15.9336, \,}\\
&{\scriptstyle 15.9336, \, 17.4380})\\ \\
\sigma=(&{\scriptstyle 0.6038, \,   0.6038, \,  -0.8553, \,
0.1480, \,   2.6346, \, -23.1985, \,
-0.0299, \,  23.0821, \,  -0.3950, \,
-0.3950, \,  -6.7305, \,}\\
&{\scriptstyle  -2.4902, \, 0.3950, \, -10.4397, \,   1.6751, \,
-0.0033, \,   0.0056, \,   0.3950, \,
7.0355, \,  -0.2005, \,   0.0056, \,
-0.1101, \,   0.2485, \,}\\  
&{\scriptstyle  2.4902, \, 4.0283}),
\end{align*}
\end{small}
where the sign pattern is 
\begin{center}
 $\text{sign}(\alpha)=\text{sign}(\sigma)=({\scriptstyle +,\,  +,\, -,\,+,\,  +,\,  -,\,-,\, +,\, -,\, -,\,  -,\,  
-,\, +,\,   -,\,  +,\,  -,\, +,\,  +,\,+,\,   -,\, +, \, -,\, +,\, +,\, +})$.
\end{center}

\medskip

We present now the matrices and vectors described in Section~\ref{sec:methods} for 
studying the capacity for multistationarity of the MAPK network without feedback.

We present below the matrices $N$ and $M$, and vector $\lambda$ 
where the order for the reactions is defined by the 
subindices of the rate constants.

\medskip

\begin{scriptsize}
$M=\left(\begin{array}{cccccccccccccccccc}
1 & 0 & 1 & 0 & 0 & 0 & 0 & 0 & 0 & 0 & 0 & 0 & 0 & 0 & 0 & 0 & 0 & 0 \\
1 & 0 & 0 & 0 & 0 & 0 & 0 & 0 & 0 & 0 & 0 & 0 & 0 & 0 & 0 & 0 & 0 & 0 \\
0 & 0 & 1 & 0 & 0 & 0 & 0 & 0 & 0 & 0 & 0 & 0 & 0 & 0 & 0 & 0 & 0 & 0 \\
0 & 1 & 1 & 0 & 0 & 0 & 0 & 0 & 0 & 0 & 0 & 0 & 0 & 0 & 0 & 0 & 0 & 0 \\
0 & 1 & 0 & 0 & 0 & 0 & 0 & 0 & 0 & 0 & 0 & 0 & 0 & 0 & 0 & 0 & 0 & 0 \\
0 & 0 & 1 & 0 & 0 & 0 & 0 & 0 & 0 & 0 & 0 & 0 & 0 & 0 & 0 & 0 & 0 & 0 \\
0 & 0 & 0 & 1 & 0 & 0 & 0 & 0 & 1 & 0 & 0 & 0 & 0 & 0 & 0 & 0 & 0 & 0 \\
0 & 0 & 0 & 1 & 0 & 0 & 0 & 0 & 0 & 0 & 0 & 0 & 0 & 0 & 0 & 0 & 0 & 0 \\
0 & 0 & 0 & 0 & 0 & 0 & 0 & 0 & 1 & 0 & 0 & 0 & 0 & 0 & 0 & 0 & 0 & 0 \\
0 & 0 & 0 & 0 & 1 & 0 & 1 & 0 & 0 & 0 & 0 & 0 & 0 & 0 & 0 & 0 & 0 & 0 \\
0 & 0 & 0 & 0 & 1 & 0 & 0 & 0 & 0 & 0 & 0 & 0 & 0 & 0 & 0 & 0 & 0 & 0 \\
0 & 0 & 0 & 0 & 0 & 0 & 1 & 0 & 0 & 0 & 0 & 0 & 0 & 0 & 0 & 0 & 0 & 0 \\
0 & 0 & 0 & 0 & 0 & 1 & 1 & 0 & 0 & 0 & 0 & 0 & 0 & 0 & 0 & 0 & 0 & 0 \\
0 & 0 & 0 & 0 & 0 & 1 & 0 & 0 & 0 & 0 & 0 & 0 & 0 & 0 & 0 & 0 & 0 & 0 \\
0 & 0 & 0 & 0 & 0 & 0 & 1 & 0 & 0 & 0 & 0 & 0 & 0 & 0 & 0 & 0 & 0 & 0 \\
0 & 0 & 0 & 0 & 0 & 0 & 0 & 1 & 1 & 0 & 0 & 0 & 0 & 0 & 0 & 0 & 0 & 0 \\
0 & 0 & 0 & 0 & 0 & 0 & 0 & 1 & 0 & 0 & 0 & 0 & 0 & 0 & 0 & 0 & 0 & 0 \\
0 & 0 & 0 & 0 & 0 & 0 & 0 & 0 & 1 & 0 & 0 & 0 & 0 & 0 & 0 & 0 & 0 & 0 \\
0 & 0 & 0 & 0 & 0 & 0 & 0 & 0 & 0 & 1 & 0 & 0 & 0 & 0 & 1 & 0 & 0 & 0 \\
0 & 0 & 0 & 0 & 0 & 0 & 0 & 0 & 0 & 1 & 0 & 0 & 0 & 0 & 0 & 0 & 0 & 0 \\
0 & 0 & 0 & 0 & 0 & 0 & 0 & 0 & 0 & 0 & 0 & 0 & 0 & 0 & 1 & 0 & 0 & 0 \\
0 & 0 & 0 & 0 & 0 & 0 & 0 & 0 & 0 & 0 & 1 & 0 & 1 & 0 & 0 & 0 & 0 & 0 \\
0 & 0 & 0 & 0 & 0 & 0 & 0 & 0 & 0 & 0 & 1 & 0 & 0 & 0 & 0 & 0 & 0 & 0 \\
0 & 0 & 0 & 0 & 0 & 0 & 0 & 0 & 0 & 0 & 0 & 0 & 1 & 0 & 0 & 0 & 0 & 0 \\
0 & 0 & 0 & 0 & 0 & 0 & 0 & 0 & 0 & 0 & 0 & 1 & 1 & 0 & 0 & 0 & 0 & 0 \\
0 & 0 & 0 & 0 & 0 & 0 & 0 & 0 & 0 & 0 & 0 & 1 & 0 & 0 & 0 & 0 & 0 & 0 \\
0 & 0 & 0 & 0 & 0 & 0 & 0 & 0 & 0 & 0 & 0 & 0 & 1 & 0 & 0 & 0 & 0 & 0 \\
0 & 0 & 0 & 0 & 0 & 0 & 0 & 0 & 0 & 0 & 0 & 0 & 0 & 1 & 1 & 0 & 0 & 0 \\
0 & 0 & 0 & 0 & 0 & 0 & 0 & 0 & 0 & 0 & 0 & 0 & 0 & 1 & 0 & 0 & 0 & 0 \\
0 & 0 & 0 & 0 & 0 & 0 & 0 & 0 & 0 & 0 & 0 & 0 & 0 & 0 & 1 & 0 & 0 & 0 \\
0 & 0 & 0 & 0 & 0 & 0 & 0 & 0 & 0 & 0 & 0 & 0 & 0 & 0 & 0 & 1 & 0 & 1 \\
0 & 0 & 0 & 0 & 0 & 0 & 0 & 0 & 0 & 0 & 0 & 0 & 0 & 0 & 0 & 1 & 0 & 0 \\
0 & 0 & 0 & 0 & 0 & 0 & 0 & 0 & 0 & 0 & 0 & 0 & 0 & 0 & 0 & 0 & 0 & 1 \\
0 & 0 & 0 & 0 & 0 & 0 & 0 & 0 & 0 & 0 & 0 & 0 & 0 & 0 & 0 & 0 & 1 & 1 \\
0 & 0 & 0 & 0 & 0 & 0 & 0 & 0 & 0 & 0 & 0 & 0 & 0 & 0 & 0 & 0 & 1 & 0 \\
0 & 0 & 0 & 0 & 0 & 0 & 0 & 0 & 0 & 0 & 0 & 0 & 0 & 0 & 0 & 0 & 0 & 1
\end{array}\right)$
\end{scriptsize}

\medskip

{\footnotesize $\lambda=0.01(1, 1, 1, 1, 1, 0.1, 0.1, 1, 1, 1, 0.001, 0.001, 0.001, 1, 0.1, 0.00001, 0.0001, 0.00001)$}.

\begin{landscape}

\noindent {\footnotesize $A=\left(\begin{array}{rrrrrrrrrrrrrrrrrrrrrrrrr}
1 & 0 &  0 &  0 &  0 &  0 &  0 &  0 & 0 &  1 & 0 &  0 & 1 &  0 &  0 & 0 &  0 & 1 &  0 &  0 &  0 & 0 & 0 & 0 & 0\\
0 & 1 & -3 & -2 & -1 &  1 &  0 & -1 & 0 & -1 & 0 &  0 & 0 & -2 & -1 & 0 & -1 & 0 & -1 & -2 & -1 & 0 & 0 & 0 & 0\\
0 & 0 &  1 &  1 &  1 & -1 &  0 &  1 & 0 &  0 & 0 &  1 & 0 &  1 &  1 & 0 &  1 & 0 &  1 &  1 &  1 & 0 & 1 & 1 & 0\\
0 & 0 & -1 & -1 & -1 &  2 &  1 &  0 & 0 &  0 & 0 &  0 & 0 & -1 & -1 & 1 &  0 & 0 & -1 & -1 &  0 & 1 & 0 & 0 & 0\\
0 & 0 &  4 &  3 &  2 & -2 & -1 &  0 & 0 &  0 & 1 &  0 & 0 &  4 &  3 & 0 &  1 & 0 &  3 &  4 &  1 & 0 & 0 & 0 & 0\\
0 & 0 &  0 &  0 &  0 &  0 &  0 &  0 & 1 &  1 & 0 &  0 & 1 &  0 &  0 & 0 &  0 & 1 &  0 &  0 &  0 & 0 & 0 & 0 & 0\\
0 & 0 &  0 &  0 &  0 &  0 &  0 &  0 & 0 &  0 & 0 & -1 & 0 &  0 &  0 & 0 &  0 & 0 &  0 &  0 &  0 & 0 & 0 & 0 & 1
   \end{array}\right)$}

\bigskip

\resizebox{\linewidth}{!}{%
$N=\left(
\begin{array}{cccccccccccccccccccccccccccccccccccc}
-1 & 1 & 0 & 0 & 0 & 1 & 0 & 0 & 0 & 0 & 0 & 0 & 0 & 0 & 0 & 0 & 0 & 0 & 0 & 0 & 0 & 0 & 0 & 0 & 0 & 0 & 0 & 0 & 0 & 0 & 0 & 0 & 0 & 0 & 0 & 0 \\
0 & 0 & 1 & -1 & 1 & 0 & -1 & 1 & 1 & -1 & 1 & 1 & 0 & 0 & 0 & 0 & 0 & 0 & 0 & 0 & 0 & 0 & 0 & 0 & 0 & 0 & 0 & 0 & 0 & 0 & -1 & 1 & 0 & 0 & 0 & 1 \\
0 & 0 & 0 & 0 & 0 & 0 & -1 & 1 & 0 & 0 & 0 & 0 & 0 & 0 & 0 & 0 & 0 & 1 & 0 & 0 & 0 & 0 & 0 & 0 & 0 & 0 & 0 & 0 & 0 & 0 & 0 & 0 & 0 & 0 & 0 & 0 \\
0 & 0 & 0 & 0 & 0 & 0 & 0 & 0 & 1 & -1 & 1 & 0 & 0 & 0 & 1 & -1 & 1 & 0 & 0 & 0 & 0 & 0 & 0 & 0 & 0 & 0 & 0 & 0 & 0 & 0 & 0 & 0 & 0 & 0 & 0 & 0 \\
0 & 0 & 0 & 0 & 0 & 0 & 0 & 0 & 0 & 0 & 0 & 1 & -1 & 1 & 0 & 0 & 0 & 0 & -1 & 1 & 1 & -1 & 1 & 1 & 0 & 0 & 0 & 0 & 0 & 0 & 0 & 0 & 0 & 0 & 0 & 0 \\
0 & 0 & 0 & 0 & 0 & 0 & 0 & 0 & 0 & 0 & 0 & 0 & 0 & 0 & 0 & 0 & 0 & 0 & -1 & 1 & 0 & 0 & 0 & 0 & 0 & 0 & 0 & 0 & 0 & 1 & 0 & 0 & 0 & 0 & 0 & 0 \\
0 & 0 & 0 & 0 & 0 & 0 & 0 & 0 & 0 & 0 & 0 & 0 & 0 & 0 & 0 & 0 & 0 & 0 & 0 & 0 & 1 & -1 & 1 & 0 & 0 & 0 & 1 & -1 & 1 & 0 & 0 & 0 & 0 & 0 & 0 & 0 \\
0 & 0 & 0 & 0 & 0 & 0 & 0 & 0 & 0 & 0 & 0 & 0 & 0 & 0 & 0 & 0 & 0 & 0 & 0 & 0 & 0 & 0 & 0 & 1 & -1 & 1 & 0 & 0 & 0 & 0 & -1 & 1 & 1 & 0 & 0 & 0 \\
-1 & 1 & 1 & 0 & 0 & 0 & 0 & 0 & 0 & 0 & 0 & 0 & 0 & 0 & 0 & 0 & 0 & 0 & 0 & 0 & 0 & 0 & 0 & 0 & 0 & 0 & 0 & 0 & 0 & 0 & 0 & 0 & 0 & 0 & 0 & 0 \\
0 & 0 & 0 & -1 & 1 & 1 & 0 & 0 & 0 & 0 & 0 & 0 & 0 & 0 & 0 & 0 & 0 & 0 & 0 & 0 & 0 & 0 & 0 & 0 & 0 & 0 & 0 & 0 & 0 & 0 & 0 & 0 & 0 & 0 & 0 & 0 \\
0 & 0 & 0 & 0 & 0 & 0 & 0 & 0 & 0 & 0 & 0 & 0 & -1 & 1 & 1 & -1 & 1 & 1 & 0 & 0 & 0 & 0 & 0 & 0 & -1 & 1 & 1 & -1 & 1 & 1 & 0 & 0 & 0 & 0 & 0 & 0 \\
0 & 0 & 0 & 0 & 0 & 0 & 0 & 0 & 0 & 0 & 0 & 0 & 0 & 0 & 0 & 0 & 0 & 0 & 0 & 0 & 0 & 0 & 0 & 0 & 0 & 0 & 0 & 0 & 0 & 0 & 0 & 0 & 0 & -1 & 1 & 1 \\
1 & -1 & -1 & 0 & 0 & 0 & 0 & 0 & 0 & 0 & 0 & 0 & 0 & 0 & 0 & 0 & 0 & 0 & 0 & 0 & 0 & 0 & 0 & 0 & 0 & 0 & 0 & 0 & 0 & 0 & 0 & 0 & 0 & 0 & 0 & 0 \\
0 & 0 & 0 & 0 & 0 & 0 & 1 & -1 & -1 & 0 & 0 & 0 & 0 & 0 & 0 & 0 & 0 & 0 & 0 & 0 & 0 & 0 & 0 & 0 & 0 & 0 & 0 & 0 & 0 & 0 & 0 & 0 & 0 & 0 & 0 & 0 \\
0 & 0 & 0 & 0 & 0 & 0 & 0 & 0 & 0 & 1 & -1 & -1 & 0 & 0 & 0 & 0 & 0 & 0 & 0 & 0 & 0 & 0 & 0 & 0 & 0 & 0 & 0 & 0 & 0 & 0 & 0 & 0 & 0 & 0 & 0 & 0 \\
0 & 0 & 0 & 0 & 0 & 0 & 0 & 0 & 0 & 0 & 0 & 0 & 0 & 0 & 0 & 0 & 0 & 0 & 1 & -1 & -1 & 0 & 0 & 0 & 0 & 0 & 0 & 0 & 0 & 0 & 0 & 0 & 0 & 0 & 0 & 0 \\
0 & 0 & 0 & 0 & 0 & 0 & 0 & 0 & 0 & 0 & 0 & 0 & 0 & 0 & 0 & 0 & 0 & 0 & 0 & 0 & 0 & 1 & -1 & -1 & 0 & 0 & 0 & 0 & 0 & 0 & 0 & 0 & 0 & 0 & 0 & 0 \\
0 & 0 & 0 & 1 & -1 & -1 & 0 & 0 & 0 & 0 & 0 & 0 & 0 & 0 & 0 & 0 & 0 & 0 & 0 & 0 & 0 & 0 & 0 & 0 & 0 & 0 & 0 & 0 & 0 & 0 & 0 & 0 & 0 & 0 & 0 & 0 \\
0 & 0 & 0 & 0 & 0 & 0 & 0 & 0 & 0 & 0 & 0 & 0 & 1 & -1 & -1 & 0 & 0 & 0 & 0 & 0 & 0 & 0 & 0 & 0 & 0 & 0 & 0 & 0 & 0 & 0 & 0 & 0 & 0 & 0 & 0 & 0 \\
0 & 0 & 0 & 0 & 0 & 0 & 0 & 0 & 0 & 0 & 0 & 0 & 0 & 0 & 0 & 1 & -1 & -1 & 0 & 0 & 0 & 0 & 0 & 0 & 0 & 0 & 0 & 0 & 0 & 0 & 0 & 0 & 0 & 0 & 0 & 0 \\
0 & 0 & 0 & 0 & 0 & 0 & 0 & 0 & 0 & 0 & 0 & 0 & 0 & 0 & 0 & 0 & 0 & 0 & 0 & 0 & 0 & 0 & 0 & 0 & 1 & -1 & -1 & 0 & 0 & 0 & 0 & 0 & 0 & 0 & 0 & 0 \\
0 & 0 & 0 & 0 & 0 & 0 & 0 & 0 & 0 & 0 & 0 & 0 & 0 & 0 & 0 & 0 & 0 & 0 & 0 & 0 & 0 & 0 & 0 & 0 & 0 & 0 & 0 & 1 & -1 & -1 & 0 & 0 & 0 & 0 & 0 & 0 \\
0 & 0 & 0 & 0 & 0 & 0 & 0 & 0 & 0 & 0 & 0 & 0 & 0 & 0 & 0 & 0 & 0 & 0 & 0 & 0 & 0 & 0 & 0 & 0 & 0 & 0 & 0 & 0 & 0 & 0 & 1 & -1 & -1 & 0 & 0 & 0 \\
0 & 0 & 0 & 0 & 0 & 0 & 0 & 0 & 0 & 0 & 0 & 0 & 0 & 0 & 0 & 0 & 0 & 0 & 0 & 0 & 0 & 0 & 0 & 0 & 0 & 0 & 0 & 0 & 0 & 0 & 0 & 0 & 0 & 1 & -1 & -1 \\
0 & 0 & 0 & 0 & 0 & 0 & 0 & 0 & 0 & 0 & 0 & 0 & 0 & 0 & 0 & 0 & 0 & 0 & 0 & 0 & 0 & 0 & 0 & 0 & 0 & 0 & 0 & 0 & 0 & 0 & 0 & 0 & 1 & -1 & 1 & 0
 \end{array}\right)$
 }
\end{landscape}

\end{document}